\RequirePackage{fix-cm}
\documentclass[smallextended]{svjour3wide} 
\smartqed  
\usepackage{amsmath,amssymb,bm}
\usepackage{mathrsfs,amsbsy,amssymb,latexsym,amsfonts,amsmath,fancyhdr}
\usepackage[dvipdfmx]{graphicx}
\usepackage{epsfig,amsfonts,amsbsy}
\usepackage{fancyhdr}
\usepackage{color}
\usepackage{url}

\newcommand{\bra}{\left\langle}
\newcommand{\ket}{\right\rangle}

\newcommand{\pder}[2]{\frac{\partial #1}{\partial  #2}}

\newcommand{\der}[2]{\frac{d #1}{d  #2}}

\newcommand{\var}[2]{\frac{\delta #1}{\delta  #2}}

\newcommand{\bv}[1]{{\boldsymbol #1}}

\newcommand{\ep}{\epsilon}

\newcommand{\DL}{D^{\rm L}}
\newcommand{\DG}{D^{\rm G}}
\newcommand{\Xeq}{X^{\rm eq}}

\newcommand{\rhoLc}{\rho^{\rm L}_{\rm c}}
\newcommand{\rhoGc}{\rho^{\rm G}_{\rm c}}
\newcommand{\muc}{\mu_{\rm c}}
\newcommand{\pc}{p_{\rm c}}

\newcommand{\sigmaL}{\sigma^{\rm L}}
\newcommand{\sigmaG}{\sigma^{\rm G}}

\newcommand{\rhoLX}{{\rho}^{\rm L}_X}
\newcommand{\rhoGX}{{\rho}^{\rm G}_X}
\newcommand{\rhoLXs}{{\rho}^{\rm L}_{X_*}}
\newcommand{\rhoGXs}{{\rho}^{\rm G}_{X_*}}

\newcommand{\barrhoLX}{{\bar\rho}^{\rm L}_X}
\newcommand{\barrhoGX}{{\bar\rho}^{\rm G}_X}
\newcommand{\barrhoLXs}{{\bar\rho}^{\rm L}_{X_*}}
\newcommand{\barrhoGXs}{{\bar\rho}^{\rm G}_{X_*}}

\newcommand{\barmuLX}{{\bar\mu}^{\rm L}_X}
\newcommand{\barmuGX}{{\bar\mu}^{\rm G}_X}
\newcommand{\barmuLXs}{{\bar\mu}^{\rm L}_{X_*}}
\newcommand{\barmuGXs}{{\bar\mu}^{\rm G}_{X_*}}

\newcommand{\pLX}{p^{\rm L}_X}
\newcommand{\pGX}{p^{\rm G}_X}
\newcommand{\pLXs}{p^{\rm L}_{X_*}}
\newcommand{\pGXs}{p^{\rm G}_{X_*}}

\newcommand{\muLX}{\mu^{\rm L}_X}
\newcommand{\muGX}{\mu^{\rm G}_X}

\newcommand{\rhoL}{\rho^{\rm L}}
\newcommand{\rhoG}{\rho^{\rm G}}

\newcommand{\eq}{{\rm eq}}
\renewcommand{\ss}{{\rm ss}}
\newcommand{\inter}{{\rm I}}

\newcommand{\Ys}{\hat Y}
\newcommand{\Yu}{X_{\uparrow}}
\newcommand{\Yd}{X_{\downarrow}}
\newcommand{\hatYu}{{\hat X}_{\uparrow}}
\newcommand{\hatYd}{{\hat X}_{\downarrow}}

\newcommand{\calF}{{\cal F}}
\newcommand{\calV}{{\cal F}_{\rm ss}}
\newcommand{\calVeq}{{\cal F}_{\rm eq}}

\renewcommand{\path}{{\rm path}}

\newcommand{\keff}{\kappa_{\Lambda}}
\newcommand{\Teff}{T_{\rm eff}}

\newcommand{\Xs}{X_*}

\newcommand{\ML}{M^{\rm L}}
\newcommand{\MLX}{M^{\rm L}_X}

\newcommand{\red}[1]{#1}

\makeatletter
\@addtoreset{equation}{section}
\makeatother
\renewcommand{\theequation}{\thesection.\arabic{equation}}

\journalname{Journal of Statistical Physics}
\begin{document}

\title{Non-equilibrium phase coexistence in boundary-driven diffusive
systems}

\author{Shin-ichi Sasa \and  Naoko Nakagawa}

\institute{
S.-i. Sasa \at
              Department of Physics, Kyoto University, Kyoto 606-8502, Japan 
              \email{sasa@scphys.kyoto-u.ac.jp}
\and 
N. Nakagawa \at
  Department of Physics, Ibaraki University, Mito 310-8512, Japan             
              \email{naoko.nakagawa.phys@vc.ibaraki.ac.jp}
}

\date{\today}

\maketitle

\begin{abstract}
Liquid-gas phase coexistence in a boundary-driven diffusive system
is studied by analyzing fluctuating hydrodynamics of a density field
defined on a
one-dimensional lattice with a space interval $\Lambda$. When an interface
width $\ell$ is much larger than $\Lambda$, the discrete model becomes the
standard fluctuating hydrodynamics, where the phase coexistence condition
is given by the local equilibrium thermodynamics. In contrast,
when $\ell < \Lambda$, the most probable density profile is determined
by a new variational principle, where the chemical potential at the
interface is found to deviate from the equilibrium coexistence chemical
potential. 
This means that metastable states at equilibrium stably
appear near the interface as the influence of the particle current.
The variational function
derived in the theoretical analysis is  also found to be equivalent
to the variational function formulated in an extended framework of
thermodynamics called global thermodynamics. Finally, the validity of the
theoretical result is confirmed by numerical simulations. 
\end{abstract}


\setcounter{tocdepth}{3}


\section{Introduction}
\label{intro}


A rich variety of phenomena exhibit non-equilibrium phase coexistence, such as boiling heat transfer, pattern formation in crystal growth, and motility-induced phase separation \cite{boiling,crystal,Cannell,Yoshida,Zhong,Ahlers,Urban,mips}. Although many such impressive phenomena are dynamic and complex, a non-trivial and surprising phenomenon has been predicted in calm and simple phase coexistence out of equilibrium. One example is the quantitative prediction that, in liquid-gas coexistence under heat conduction, the temperature of the liquid-gas interface is lower than the equilibrium coexistence temperature for the pressure \cite{Global-PRL,Global-JSP}, where  the equilibrium phase coexistence occurs at the
first-order phase transition point far from the critical point. This phenomenon
means that metastable states at equilibrium are stabilized by a steady
current even in the linear response regime.  


The prediction was presented in an extended framework of thermodynamics that we call {\it global thermodynamics}. This framework was first proposed as a natural extension of the minimum principle of free energy with the key concept of global temperature \cite{Global-PRL}. Applying the framework to a van-der Waals fluid revealed that the temperature of the liquid-gas interface is different from the first-order transition temperature at equilibrium. Then, the formulation was carefully arranged so that quantitative predictions could be made for real materials \cite{Global-JSP}. The equivalence among different ensembles was discussed, and finally, the maximum entropy principle was formulated for enthalpy-conserving heat conduction systems \cite{Global-PRR}. The entropy defined in the formulation is found to possess a non-additive term in addition to the space integral of the local entropy density field. This formulation enables us to apply global thermodynamics to various systems. 


There have been no experimental reports on the predictions of global thermodynamics. Nevertheless, it is worth mentioning that numerical simulation of the Hamiltonian Potts model in heat conduction shows results consistent with the quantitative prediction of global thermodynamics for large enough systems \cite{KNS}. The singular nature of the phase coexistence has also been discussed by analyzing a mesoscopic model describing the order parameter dynamics in heat conduction \cite{Global-PRE}. However, this analysis involves some phenomenological assumptions in calculating the singular part. Furthermore, the model is too complicated to extract the microscopic mechanism of the deviation in interface temperature from the equilibrium phase coexistence temperature.


On the basis of this background, we study a simple system that exhibits non-equilibrium phase coexistence. We consider a system in which the number density field is a single dynamical variable and the density field is directly driven by a boundary condition of the chemical potential, where the temperature is given as a constant parameter of the system. The stochastic time evolution of the density field is described in terms of a discrete form of fluctuating hydrodynamics with
a space interval $\Lambda$. When the width of an interface in  phase coexistence $\ell$ is much larger than $\Lambda$, the model is equivalent to the standard fluctuating hydrodynamics \cite{Schmitz,Eyink,Spohn,Bertini-RMP}. Then, local fluctuation properties of thermodynamic quantities are described by local equilibrium distribution  \cite{Kuramoto}. In contrast, when $\ell \ll \Lambda$, the local distribution may be out of equilibrium \cite{Yana}.  For this case, 
we derive the variational function that determines the most probable density profile. We then find that the chemical potential at the interface
of the most probable profile deviates from the equilibrium coexistence
value. This means that metastable states at equilibrium stably appear
near the interface of the driven system.
The formula describing the deviation takes the same
form as those phenomenologically predicted by global thermodynamics.
Indeed, we can \red{derive the variational function
for this system by using the method of global thermodynamics.} 
We also confirm the validity of the theoretical calculation by numerical
simulations.


The rest of this paper is organized as follows.
In Sec. \ref{setup}, we introduce a \red{stochastic} 
model we study in this paper.
We then review phase coexistence conditions for equilibrium systems,
and summarizes basic issues for non-equilibrium phase coexistence.
In Sec. \ref{Sd}, we derive a variational function by analyzing
the Zubarev-McLennan representation of the stationary distribution.
In Sec. \ref{Val}, we rewrite the variational equation as the form
giving the chemical potential at the interface. 
In Sec. \ref{Global}, we \red{derive the variational function
using the method of global thermodynamics.} 
In Sec. \ref{ns}, we show results of
numerical simulations. Section \ref{concluding} provides some
concluding remarks.

\section{Setup}
\label{setup}

\subsection{Model}

\begin{figure}[t]
\centering
\includegraphics[scale=0.5]{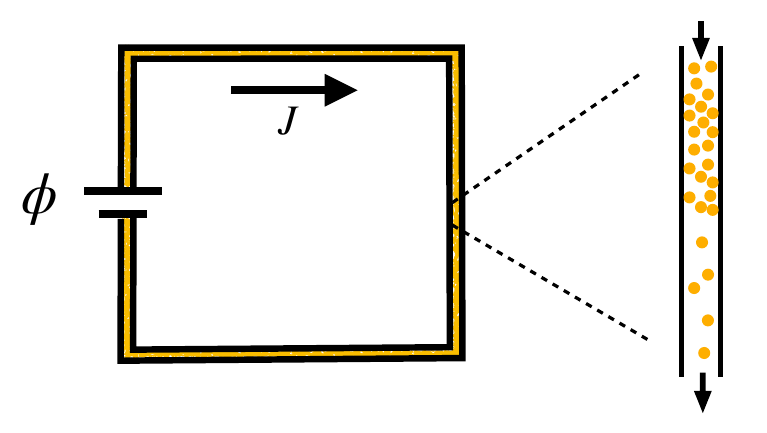}
\caption{
  Schematics of a system we study. A one-dimensional density field is
  defined on a  ring with a battery at $x=0$.}
\label{fig:setup}
\end{figure}

We consider a collection of stochastic and diffusive particles in a closed tube
which are driven by an external battery at one surface $x=0$. See
Fig. \ref{fig:setup} for the illustration of the setup.
We describe
the system by an averaged particle density $\rho(x)$ defined in a
one-dimensional region $[0,L]$. More precisely, we define
\begin{equation}
  \rho(x) \equiv \frac{1}{A}\int dy dz~  \rho(x,y,z)
\end{equation}  
for the three-dimensional particle density $\rho(x,y,z)$,
where the area of the cross section of the tube is denoted by $A$. 
We then assume a standard continuum description of fluctuating hydrodynamics
of $\rho(x)$. The density field $\rho(x,t)$ satisfies the continuity 
equation
\begin{equation}
  \partial_t \rho+ \partial_x j=0.
\label{continuity}
\end{equation}
Based on the mean field picture in the cross section,
we assume the current as 
\begin{equation}
  j(x,t) = -\sigma(\rho(x))\left[\partial_x
  \var{\calF}{\rho(x)}-\phi \delta(x)\right]
  +\sqrt{\frac{2\sigma(\rho(x))T}{A}}\cdot \xi(x,t).
\label{current}
\end{equation}
$T$ is the temperature, $\sigma(\rho)$ is a conductivity
as a smooth function of $\rho$, and
$\phi$ represents the voltage of a battery located at $x=0$.
$\xi$ is Gaussian-white noise satisfying $\bra \xi \ket=0$ and
\begin{equation}
  \bra \xi(x,t) \xi(x',t') \ket
  =\delta(x-x')\delta(t-t').
\end{equation}
The symbol ``$\cdot$" in front of $\xi$ in (\ref{current})
represents the Stratonovich product in space and the Ito product in time.
Here, because the variance of the surface average 
$\Xi(x,t) \equiv \int dy dz \xi(x,y,z,t)/A$
  of the three-dimensional  Gaussian-white noise $\xi(x,y,z,t)$
with unit variance
is $1/A$, we set $\Xi(x,t)=\xi(x,t)/\sqrt{A}$, which determines
the $A$ dependence of \eqref{current}.
The free energy functional $\calF$ of the density profile
$\bv{\rho}=(\rho(x))_{0 \le x \le L}$  
is expressed as 
\begin{equation}
  \calF(\bv{\rho})=\int_0^L dx \left[
    f(\rho(x))+ \frac{\kappa}{2} (\partial_x \rho)^2
    \right].
\label{Free-form-con}  
\end{equation}
As a specific example, we consider the case that 
\begin{equation}
  f(\rho)= -\frac{1}{2}(\rho-1.5)^2+\frac{1}{4}(\rho-1.5)^4
\label{free}  
\end{equation}
by introducing dimensionless length and energy in this form. 
Note that our  argument below is independent of
the specific form if $f(\rho)$ contains two local minima.
The length unit is assumed to be the order of particle
distances. $\kappa$ characterizes the interface free energy, which
is relevant when $\partial_x \rho$ is large. In particular,
when $0.5 < \bar \rho < 2.5$ in noiseless equilibrium
systems with $T=0$ and $\phi=0$, phase coexistence occurs
with two interfaces. Then, $\sqrt{\kappa}$ determines
the interface width in the phase coexistence.

Now, we notice that there  
is a cutoff length $\Lambda$ of the continuum description.
Because the noise is assumed to be white in space, 
$\Lambda$ should be larger than the microscopic length,
which is set to the order of unity.
On the other hand, it is obvious that $\Lambda$ should be 
much smaller than the system size $L$. In many cases, the calculation
result of fluctuating hydrodynamics is independent of the cutoff
$\Lambda$ even in the limit $\Lambda \to 0$, while there is a case
where a singular cutoff dependence is observed \cite{Kado}.
Here, let us recall that the width of interfaces in phase coexistence
is estimated as a microscopic length. Thus, this may be smaller than
the cutoff length of fluctuating hydrodynamics. Such a case cannot
be studied by the continuum model. We thus need to propose and analyze  
a discrete model in which the microscopic cutoff $\Lambda$ is
explicitly introduced. 

With this background, we consider a sequence of $N$-boxes
in a one-dimensional ring. 
Let $\rho_i$ be the density of particles at the $i$-th box,
where $1 \le i \le N$. Mathematically, $\rho_i$ is defined
on the $i$-th site in the one-dimensional lattice
$\{ i | 1 \le i \le N, \ i \in \mathbb{Z} \}$  with 
the periodic boundary condition $\rho_0 \equiv \rho_N$
and $\rho_{N+1}=\rho_1$.  The horizontal size of the box is denoted
by $\Lambda$ and the cross section area of the box is $A$. 
The system size $L$ is given by $L=\Lambda N$.
The free energy functional given in (\ref{Free-form-con})
is then replaced with
\begin{equation}
  \calF(\bv{\rho})=\Lambda \sum_{i=1}^N \left[
    f(\rho_i)+ \frac{\kappa}{2\Lambda^2} (\rho_{i+1}-\rho_i)^2 
    \right]
\label{Free-form}  
\end{equation}
for $\bv{\rho}=(\rho_i)_{1 \le i \le N}$.
The density $\rho_i$ satisfies the continuity equation
\begin{equation}
  \der{\rho_i}{t}+ \frac{j_i-j_{i-1}}{\Lambda}=0
\label{model:con}  
\end{equation}
with $j_0=j_N$. The current $j_i$ is defined on the
$i$-th bond connecting the $i$-th site and the $(i+1)$-th site.
Using the generalized
chemical potential $\tilde \mu_i$ defined by
\begin{equation}
  \tilde \mu_i \equiv \frac{1}{\Lambda}\pder{{\cal F}}{\rho_i},
  \label{tilde-mu}
\end{equation}
we replace the current given in (\ref{current}) by 
\begin{equation}
  j_i(t)= - \frac{\sigma(\rho_i^{\rm m})}{\Lambda}
  (\tilde \mu_{i+1}- \tilde \mu_i - \phi \delta_{i,N}) +
  \sqrt{ \frac{2 \sigma(\rho_i^{\rm m}) T}{A \Lambda} }\cdot \xi_i(t),
\label{model:cur}
\end{equation}
where we set $\rho^{\rm m}_i= (\rho_i+\rho_{i+1})/2$ to satisfy the
detailed balance condition. $\xi_i(t)$ is 
Gaussian-white noise satisfying $\bra \xi_i \ket=0$ and
\begin{equation}
  \bra \xi_i(t) \xi_j(t') \ket
  =\delta_{ij}\delta(t-t').
\end{equation}
The $\Lambda$ dependence of the noise intensity in 
  \eqref{model:cur} is understood from the replacement of
  $\delta(x-x')$ with $\delta_{ij}/\Lambda$.
Explicitly, $\tilde \mu_i$ is written as
\begin{equation}
  \tilde \mu_i = \mu(\rho_i)-
  \frac{\kappa}{\Lambda^2}(\rho_{i+1}+\rho_{i-1}-2\rho_i)
\label{tilde-mu-2}
\end{equation}  
with  $\mu(\rho_i)= \partial f(\rho_i)/\partial \rho_i$.
The total number of particles $\sum_{i=1}^N \rho_i =\bar \rho N$
is conserved in the time evolution. The average density $\bar \rho$
is a parameter of the system. 

\begin{figure}[t]
\centering
\includegraphics[scale=0.45]{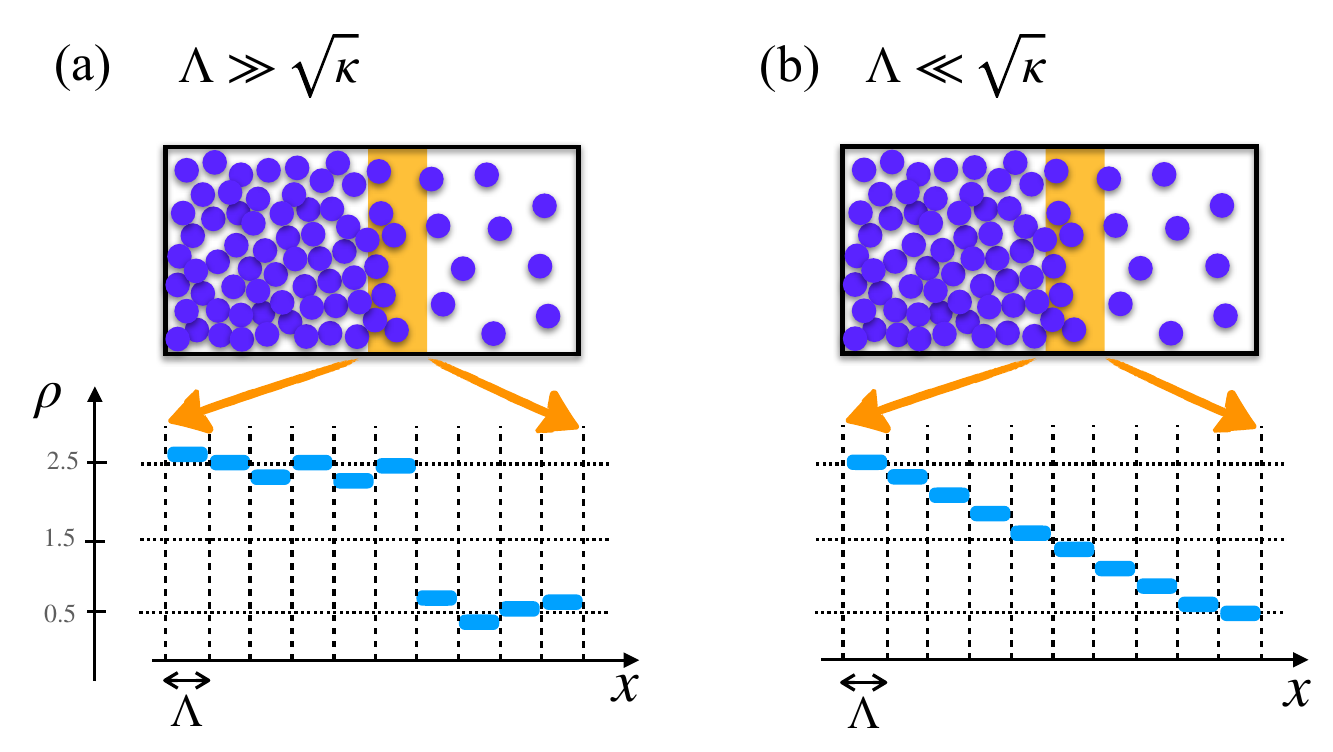}
\caption{
Illustration of two limiting cases  (a) $\keff \ll 1$ and (b) $\keff \gg 1$.}
\label{fig:Lambda}
\end{figure}

The steady state of this system is characterized by five parameters
$(\keff, \Teff, \phi, \bar \rho, N)$, where 
\begin{equation}
  \keff \equiv \frac{\kappa}{\Lambda^2}, \qquad
  \Teff\equiv \frac{T}{A}.
\label{five}
\end{equation}  
That is, systems with the same values of $(\keff, \Teff, \phi, \bar \rho, N)$
exhibit the same steady state.  In the argument below, the
$\Lambda$ dependence appear only through $\keff$ dependence.
When $\keff \gg 1$ and $N \to \infty$,
the system behavior of (\ref{model:con}) and (\ref{model:cur})
is understood by analyzing (\ref{continuity}) and (\ref{current})
because  (\ref{model:con}) and (\ref{model:cur}) correspond to
an accurate approximation of (\ref{continuity}) and (\ref{current}).
In contrast, 
the system behavior for the case $\keff < 1$ cannot be
described by (\ref{continuity}) and (\ref{current}).
For such cases, we have to analyze the discrete model with focusing 
on the limiting case that $\keff \ll 1$.  See Fig.  \ref{fig:Lambda}
for the illustration of two limiting cases.
Finally, we note that 
$A \gg 1$ because $A$ is the square of
a macroscopic length.  We thus consider the weak noise limit
$\Teff \to 0$ for the steady state realized in the limit $t \to \infty$.  

\subsection{Review of equilibrium phase coexistence}
\label{review-eq}

We review phase coexistence states for the equilibrium system
with $\phi=0$. 
The stationary distribution of $\bv{\rho}$ is given by
\begin{equation}
  {\cal P}_{\eq}(\bv{\rho};\bar \rho)=\frac{1}{Z} \exp
  \left[ -\frac{1}{\Teff} {\cal F}(\bv{\rho}) \right]
  \delta\left( \sum_{i=1}^N\rho_i-\bar \rho N \right).
\label{canonical}
\end{equation}
See Appendix \ref{app:psi} for the derivation. 
The most probable profile in the weak noise limit $\Teff \to 0$ is
determined as the density profile that minimizes ${\cal F}(\bv{\rho})$
defined by \eqref{Free-form}.
The variational equation is obtained as
\begin{equation}
\tilde\mu_i=\tilde\mu_{i+1},
\label{v-eq:eq}
\end{equation}
which means that $\tilde\mu_i$ is a constant independent of $i$.

\begin{figure}[t]
\centering
\includegraphics[scale=0.55]{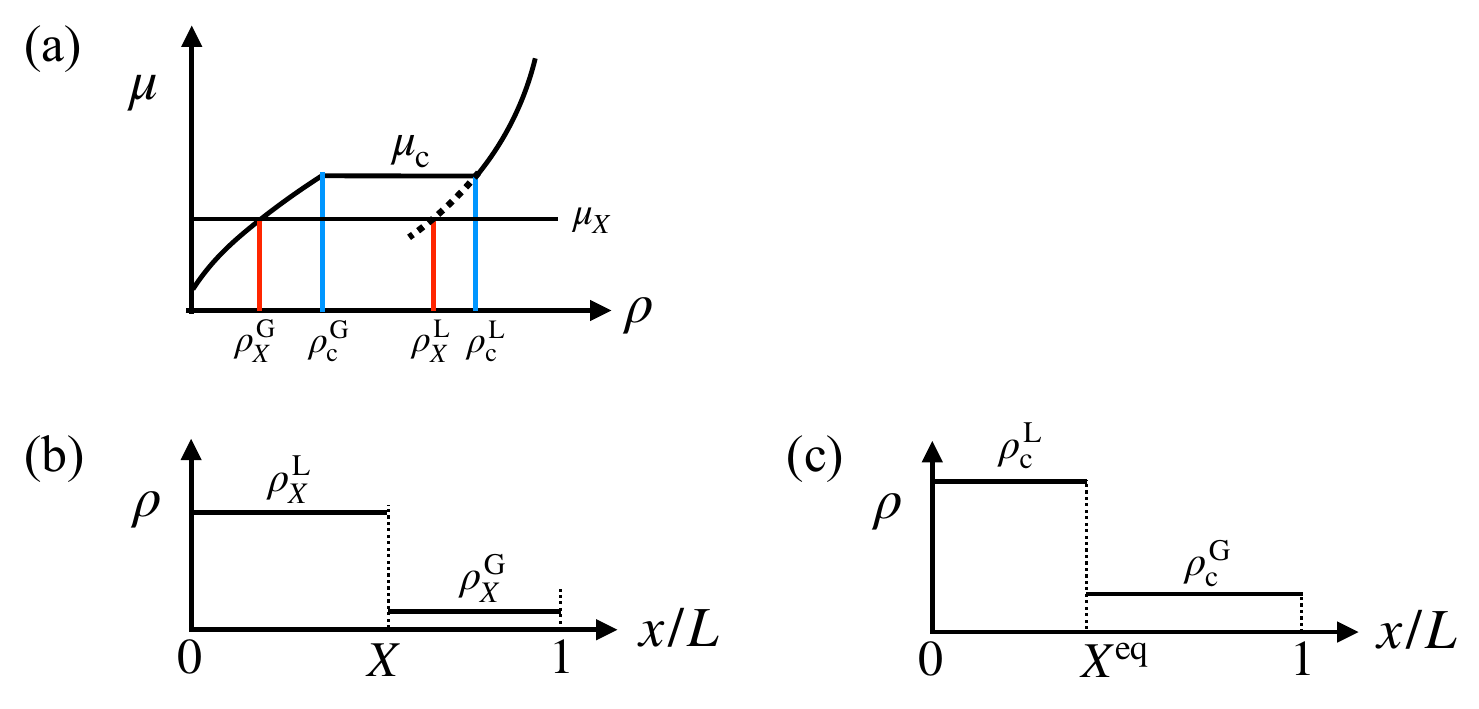}
\caption{
Illustration of metastable profile and equilibrium profile.}
\label{fig:metasta}
\end{figure}

We study two limiting cases $\keff \gg 1$ and $\keff  \ll 1$.
First, for the case $\keff \gg 1$, 
we can derive the phase coexistence condition from the analysis
of the continuum limit of (\ref{v-eq:eq}) with $N \to \infty$,
which is given in (\ref{1st}). Importantly, the solution of the
variational equation is unique under the boundary condition that 
$\partial_x \rho(0)=\partial_x\rho(L)=0$ with $\rho(0) > \rho(L)$.
As shown in Appendix \ref{app:con}, phase coexistence occurs when
$\bar \rho$  satisfies $\rhoGc \le \bar \rho \le \rhoLc$,
where $\rhoLc$ and $\rhoGc$ are determined by
\begin{equation}
\mu(\rhoLc)= \mu(\rhoGc), \qquad p(\rhoLc)= p(\rhoGc)
\label{n-con}
\end{equation}
with the thermodynamic pressure $p(\rho)$ defined by
\begin{equation}
 p(\rho)\equiv \rho \mu(\rho)-f(\rho).
\label{con-coex}
\end{equation}
Hereafter, the equilibrium value of the coexistence chemical potential
is denoted by $\mu_c\equiv \mu(\rhoLc)= \mu(\rhoGc)$.
We assume $\rhoGc < \rhoLc $ without loss of generality.
Then, $\rhoLc$ and $\rhoGc$ represent the densities of the liquid
and gas in the phase coexistence, respectively. 
Furthermore, once $\rhoLc$ and $\rhoGc$
  are obtained, the fraction of the liquid region $X^{\rm eq}$ is uniquely
  determined from $\Lambda \sum_i \rho_i=\bar \rho L$,
which is expressed by
\begin{equation}
\rhoLc \Xeq+ \rhoGc (1-\Xeq)=\bar \rho.
\label{Xeq-det}
\end{equation}
See Fig.~\ref{fig:metasta}(c) for the profile.

For the other case $ \keff \ll 1$, which we mainly study in this paper, 
the derivation method in the continuum limit cannot be applied. However,
even in this case, the variational principle determines the most probable
profile.
  The variational equation \eqref{v-eq:eq} is rewritten as $\mu_i=\mu_{i+1}$
  in the limit $\keff\to 0$ so that the chemical potential is uniform.
  In this case, there is a one-parameter family of solutions
    $\rho_{i;X}^{\phi=0}$ characterized 
    by \red{$X \in [0.1]$, where $NX$ is an half integer.}
    \red{Let us}
    write the solution $\rho_{i;X}^{\phi=0}$ explicitly.
    Referring to Fig. \ref{fig:metasta}(a), 
  \red{for a given $X$, 
  we set $\rhoLX$ and $\rhoGX$
  as those satisfying 
\begin{equation}
 \mu(\rhoLX)=\mu(\rhoGX), \qquad X \rhoLX+ (1-X)\rhoGX =\bar \rho 
\label{mueq}
\end{equation}
with $\rhoGX < \rhoLX$. }
We can then express the solution as
\begin{equation}
  \rho_{i;X}^{\phi=0}=\rhoLX \chi(i \in [1, NX])+
  \rhoGX \chi(i \in [NX,N]),
\label{rhoXeq}  
\end{equation}
where $\chi( P )$ represents the
characteristic function that takes $1$ if $P$ is true
and $0$ otherwise.  \red{From Fig. \ref{fig:metasta}(b),  
it is found that $X$ corresponds to the interface position
of $\rho_{i;X}^{\phi=0}$.} 
\red{Note that the free energy of the solutions is smaller 
than the other solutions with more interfaces as far as $\keff >0$.}
\red{The uniform value of the chemical potential for the 
solution $\rho_{i;X}^{\phi=0}$ is denoted by $\mu_X^{\phi=0}$,
which is equal to $\mu(\rhoLX)$ and $\mu(\rhoGX)$.}

Note that the value of $X$ is not determined
from the variational equation (\ref{v-eq:eq}) with $\keff \to 0$.
Physically, the solutions $\bv{\rho}^{\phi=0}_X$ form a
family of metastable states characterized by $X$.
The most probable value of $X$,  which is denoted by $X_*$,
is derived from the minimum free energy principle formulated
as follows. We first define the variational function
\begin{equation}
\calVeq(X;\bar \rho)\equiv  \calF(\bv{\rho}_X^{\phi=0}).
\label{eq-var}
\end{equation}
Then, the minimum free energy principle means
\begin{equation}
  \calVeq(X_*;\bar \rho) =\min_X \calVeq(X;\bar \rho),
\label{F-min}  
\end{equation}
where, in the present case, $X_*$ satisfies
\begin{equation}
\left. \der{\calVeq(X;\bar\rho)}{X} \right|_{X=X_*}=0.
\label{var-FX-eq}
\end{equation}
Substituting the expression
\begin{equation}
  \calF(\bv{\rho}_X^{\phi=0})=LXf(\rhoLX)+L(1-X)f(\rhoGX)
\end{equation}
into (\ref{var-FX-eq}), we obtain 
\begin{equation}
f(\rhoLXs)-f(\rhoGXs)=\mu_{X_*}^{\phi=0} (\rhoLXs-\rhoGXs)
\label{var-FX*}
\end{equation}
using (\ref{mueq}) and (\ref{mueq}).
Comparing \eqref{mueq} and \eqref{var-FX*} with \eqref{n-con}, we find
\begin{equation}
\mu_c=\mu_{X_*}^{\phi=0}, \quad \rhoLc=\rhoLXs,  \qquad \rhoGc=\rhoGXs.
\label{rhoc-det}  
\end{equation}
Moreover, $X_*$ is equal to $X^{\rm eq}$. 
The result means that the equilibrium states for the two
limiting cases, $\keff \gg 1 $ and $\keff \ll 1$, are equivalent.

\subsection{Preliminaries for non-equilibrium phase coexistence}
\label{prel-ne}

We concentrate on the case that $\bar \rho$ satisfies
$\rhoGc < \bar \rho <\rhoLc$ where liquid and gas coexist.
When the voltage of the battery $\phi$ is positive, the stationary
distribution is not written as the canonical form \eqref{canonical}.
Therefore, we do not have a general variational principle for determining
the most probable profile for $\phi >0$. 
Nevertheless, we divide the problem for determining the most probable
profile into two steps. As the first step, we consider stationary solutions
of the deterministic equations without noise. If the stationary solution is
unique, this is the most probable profile in the weak noise limit. In contrast,
if stationary solutions form a one parameter family, we proceed to the second
step, where we will formulate a variational principle for selecting the most
probable profile among the stationary solutions as we have examined
in \eqref{F-min} for the  equilibrium system. In this section, we focus
on the first step.

We consider stationary solutions of the deterministic
    equation given in (\ref{continuity}) and (\ref{current}) with
    $\Teff=0$. Let $J$ be the steady particle current induced by
   the battery. Setting $j_i(t)=J$ for all $i$ in (\ref{current}), 
   we obtain the conduction equation
\begin{equation}
  \frac{1}{\Lambda}
\left(\tilde \mu_{i+1}- \tilde \mu_i + \phi \delta_{i,N} \right)
  =
-\frac{J}{\sigma(\rho_i^{\rm m})} .
\label{v-eq:neq}
\end{equation}  
The uniformity of $J$ yields
\begin{equation}
J = \phi \left( \Lambda \sum_{i=1}^N \frac{1}{\sigma(\rho_i^{\rm m})} \right)^{-1}.
\label{Jdef-dis}
\end{equation}

When $\keff \gg 1$ and $N \to \infty$ \red{with $L=N \Lambda$ and
$\kappa=\kappa_\Lambda \Lambda^2$ fixed}, we can analyze the continuum
limit of (\ref{v-eq:neq}).  As shown in Appendix \ref{app:con}, 
we find that the solution of the deterministic equation is
unique and that the chemical potential and pressure are
continuous at the interface in the space $\hat x=x/L$ with $L \to
\infty$. We can then conclude that the chemical potential
at the interface is $\muc$. See Appendix \ref{app:con} for the derivation
of this result. 

In contrast, when \red{we first fix $\keff \ll 1$ and take the
limit $N \to \infty$}, 
there is a family of stationary solutions of  the conduction 
equation (\ref{v-eq:neq}). 
In the equilibrium system, they correspond to metastable states
characterized by \eqref{mueq}, as shown in Fig. \ref{fig:metasta}(b),
and the metastable states were candidates for the most probable
profile. We thus expect that a family of stationary solutions
$\rho_{i;X}^{\phi}$ for $\phi>0$ correspond to metastable states
among which the most probable profile is selected. 
In the remainder of this section, we express the metastable
states explicitly as a preliminary for the analysis in the next section.

We concentrate on small $\phi$ and ignore the contribution of $O(\phi^2)$.
The solution $\rho_{i;X}^{\phi}$ should be given as a perturbation of
the equilibrium solution $\rho_{i;X}^{\phi=0}$ given in \eqref{rhoXeq}.
We thus assume that 
$|\rho_{i;X}^\phi-\rhoLc| < |\rho_{i;X}^\phi-\rhoGc|$ for $1 \le i < NX$,
and $|\rho_{i;X}^\phi-\rhoGc| < |\rho_{i;X}^\phi-\rhoLc|$ for $NX <  i \le N$.
Letting
\begin{equation}
\sigma^L = \sigma(\rhoLc), \qquad  \sigma^G = \sigma(\rhoGc),
\label{sigmaLG}
\end{equation}
the conduction equation \eqref{v-eq:neq} with $\keff\to 0$ results in
\begin{align}
   &\mu(\rho_{i;X}^\phi)= -
    \frac{1}{\sigma^L}JL\frac{i-1}{N}+\mu_{1:X}^\phi +O(\phi^2) , \qquad (1\le i\le NX),
\label{muLd}\\
    &\mu(\rho_{i;X}^\phi)= -\frac{1}{\sigma^G}JL\frac{i-N}{N}+\mu_{N;X}^\phi +O(\phi^2), \qquad (NX< i\le N),
    \end{align}  
where
  \begin{equation}
    \mu_{1;X}^\phi=\mu_{N;X}^\phi+\phi .
\label{muGd}    
\end{equation}
From the direct calculation, we also obtain the
  difference between the chemical potentials of two adjacent sites
  over the interface as 
  \begin{equation}
\mu(\rho_{X-1/(2N);X}^\phi)-\mu(\rho_{X+1/(2N);X}^\phi)=O(\phi/N).
\label{c-diff}
  \end{equation}
Finally, the condition 
\begin{equation}
\frac{1}{N} \sum_{i=1}^N  \rho_{i;X}^\phi = \bar \rho
\label{den-cons:d}
\end{equation}
uniquely determines $\mu_{1;X}^\phi$ and $\rho_{1;X}^\phi$
for a given $X$. We can easily confirm that
$\lim_{\phi \to 0}\rho_{i;X}^{\phi}$ is equivalent to (\ref{rhoXeq}).

\begin{figure}[t]
\centering
\includegraphics[scale=0.5]{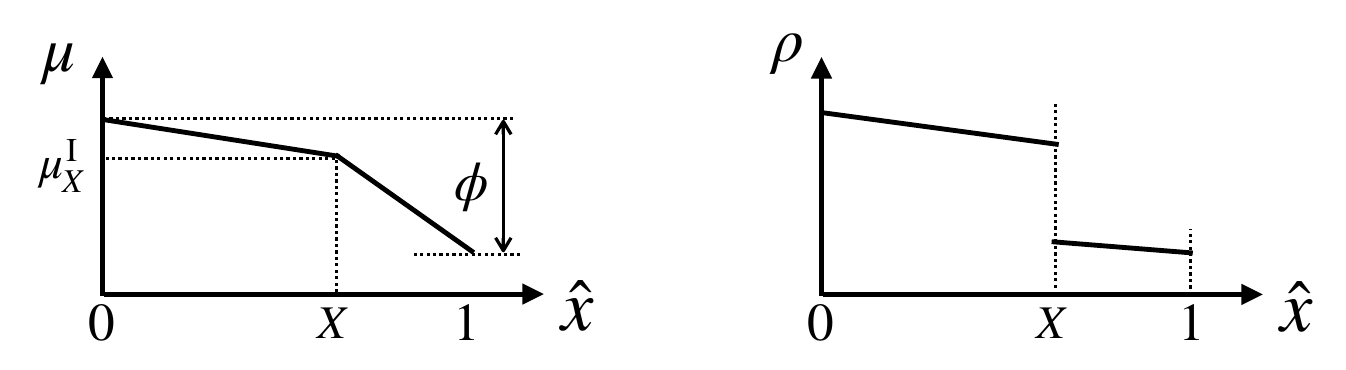}
\caption{
  Schematic profiles of $\mu_X^\phi(\hat x)$ and $\rho_X^\phi(\hat x)$.
}
\label{fig:profile}
\end{figure}

\subsection{Singular continuum description}
\label{SCD}

We introduce a continuum description
using a real variable $\hat x=x/L$ defined on the interval $[0,1]$ by taking
the limit $N \to \infty$.  The mathematical \red{formulation} 
of this limit
is beyond the scope of the present paper. However, to simplify the
calculation in the subsequent sections, we naively introduce the continuum
description based on expected behaviors of $\rho_i(t)$ for large $N$.
For a discrete variable 
$\bv{\rho}=(\rho_i(t))_{i=1}^N$ at 
time $t$, we first define $\rho_N(\hat x,\hat t)$ as the piece-wise
linear function
obtained by connecting two consecutive points $(i/N, \rho_i(t))$
and $((i+1)/N, \rho_{i+1}(t))$ for $0 \le i \le N$
in the $(\hat x, \rho)$
space, where we set $\hat t=t/L^2$ for later convenience.
For sufficiently large $N$, we expect that there exists an
almost continuous function $\rho(\hat x, \hat t)$ such that
$|\rho(\hat x, \hat t)-\rho_N(\hat x, \hat t)| =O(1/N)$. The chemical potential
$\mu(\hat x,\hat t)$ in the continuum description is defined  from
$\mu_i(t)$ by the same procedure and it is expected that
$\mu(\hat x,\hat t)=\mu(\rho(\hat x,\hat t))$. The definition of
$j(\hat x,\hat t)$
is slightly different from $\rho(\hat x,\hat t)$ and  
$\mu(\hat x, \hat t)$,
because the current $j_i(t)$ is defined on the $i$-th bond connecting the
$i$ site and $i+1$ site as seen in (\ref{model:con}) and (\ref{model:cur}).
Concretely, we define $j(\hat x,\hat t)$ by using  the piece-wise linear
function obtained by connecting points $(i/N+1/2N, \red{j_i/L})$ in the
$(\hat x, j)$ space.  Then, the continuum limit of 
\eqref{model:con}  and \eqref{model:cur} with $\keff \to 0$ is expressed by 
\begin{equation}
  \partial_{\hat t} \rho(\hat x, \hat t)+ \partial_{\hat x} j(\hat x, \hat t)
  =0,
\label{continuity-hat}
\end{equation}
and 
\begin{align}
  j(\hat x, \hat t)=-\sigma(\rho(\hat x,\hat t))
  \left[\pder{\mu(\hat x,\hat t)}{\hat x}-\phi \delta(\hat x)\right]
  +\sqrt{\frac{2\sigma(\rho(\hat x,\hat t))\Teff}{L}}\cdot
  \hat \xi(\hat x,\hat t),
\label{j-continuous}
\end{align}
where $\hat \xi(\hat x,\hat t)$ satisfies  
\begin{equation}
  \bra \hat \xi(\hat x,\hat t) \xi(\hat x',\hat t') \ket
  =\delta(\hat x-\hat x')\delta(\hat t-\hat t').
\end{equation}

Apparently, (\ref{continuity-hat}) and  
(\ref{j-continuous}), which we call {\it the singular continuum description},
take the same form as a singular case of the standard continuum description
(\ref{continuity}) and (\ref{current}) with $\kappa \to 0$ by setting
$\hat x =x/L$ and $\hat t = t/L^2$.
We here notice \red{the difference between the two descriptions.
In the singular continuum description, we first take 
$N \to \infty$  with $L$ and $\kappa_\Lambda$ fixed,
and then consider $\kappa_\Lambda \to 0$.
On the other hand, as described in the previous subsection,
in the standard continuum description, we take
$\kappa_\Lambda=\kappa/\Lambda^2 \to \infty$ and 
$N \to \infty$ with $L$ and $\kappa$ fixed.} 
Then, as a singular case of the standard continuum description,
we consider the limit $\kappa  \to 0$.
The behavior of the two descriptions are rather different, but we do not
discuss the difference anymore. In the argument below, we will consider only
the singular continuum description.

The stationary solution of the deterministic equation $\rho_{i;X}^\phi$ and the
corresponding chemical potential $\mu(\rho_{i;X}^\phi) $ are expressed
as $\rho_X^\phi(\hat x)$ and $\mu_X^\phi(\hat x)$ in the singular continuum
description. Note that 
$\rho_X^\phi(\hat x)$ is discontinuous at $\hat x=X$, while
$\mu_X^\phi(\hat x)$ is continuous at $\hat x=X$ as is seen from \eqref{c-diff}.
The chemical potential at the interface $\hat x=X$ is denoted by
\begin{equation}
  \mu_X^{\rm I} \equiv  \mu_X^\phi(X).
\label{mu-int-def}
\end{equation}  
As shown in Fig. \ref{fig:profile}, $\rho_X^\phi(\hat x)$ is 
a piece-wise continuous function and that $\mu_X^\phi(\hat x)$
is a piece-wise smooth function with singular points at $\hat x=0$
and $\hat x=X$. Furthermore, 
we rewrite (\ref{muLd}) and (\ref{muGd}) as
\begin{equation}
    \mu_X^\phi(\hat x)= -\frac{1}{\sigma^L}JL\hat x+\mu_X^\phi(0) +O(\phi^2)
    \label{muLc}
  \end{equation}
for $0 \le \hat x < X$, and 
  \begin{equation}
    \mu_X^\phi(\hat x)= -\frac{1}{\sigma^G}JL(\hat x-1)+\mu_X^\phi(1) +O(\phi^2)
\label{muGc}    
  \end{equation}
for $X <  \hat x \le 1$. We also have
  \begin{equation}
    \mu_X^\phi(0)=\mu_X^\phi(1)+\phi .
\label{muGLc}    
  \end{equation}
Here, from \eqref{Jdef-dis}, we obtain
\begin{equation}
JL= \phi \left( \frac{X}{\sigmaL}+\frac{1-X}{\sigmaG} \right)^{-1}.
\label{J-phi}
\end{equation}
Then, the density field $\rho_X^\phi(\hat x)$ for $0 \le \hat x < X$ 
is determined from (\ref{muLc}) with
$\mu_X^\phi(\hat x)=\mu(\rho_X^\phi(\hat x))$
and $|\rho_X^\phi(\hat x)-\rhoLc| < |\rho_X^\phi(\hat x)-\rhoGc|$. Similarly,
the density field $\rho_X^\phi(\hat x)$ for $X <  \hat x \le 1$ 
is determined from (\ref{muGc}) with
$\mu_X^\phi(\hat x)=\mu(\rho_X^\phi(\hat x))$
and $|\rho_X^\phi(\hat x)-\rhoGc| < |\rho_X^\phi(\hat x)-\rhoLc|$.
Note that $\rho_X^\phi(\hat x)$ and $\partial_x \mu_X^\phi(\hat x)$
are discontinuous at
$\hat x=0$ and $\hat x=X$. For given $X$ and system parameters 
$(\phi, \bar \rho)$, $\mu_X^\phi(\hat x)$ and $\rho_X^\phi(\hat x)$ are 
uniquely determined  from the condition
\begin{equation}
  \int_0^1 d\hat x~ \rho_X^\phi(\hat x) =\bar \rho .
\label{barrho-def}  
\end{equation}

\section{Variational function}
\label{Sd}

To simplify the notation, we use $x$ and $t$ for $\hat x$ and $\hat t$
in this and next sections. 
In the previous section, we determined the candidates of steady density
profile $\rho_X^\phi(x)$ characterized by $X$ in the weak noise
limit $\Teff \to 0$. To determine the  most probable density profile
among them for  given system parameters $(\bar\rho,\phi)$,
in Sec. \ref{sec:zm},
we derive a variational function using the Zubarev-McLennan 
representation of the steady state.  The variational function includes
a time integral of the current at $x=0$. After confirming some basic
issues and assumptions in Sec. \ref{Cal-Q}, we calculate the time
integral of the current in Sec. \ref{cal-j}. The result is presented in
Sec. \ref{result-j}.

\subsection{Stationary distribution}
\label{sec:zm}

We consider the stationary distribution
${\cal P}_\ss(\bv{\rho};\bar \rho, \phi)$ 
of density field $\bv{\rho}$.
When $\phi=0$, the stationary distribution is given by
(\ref{canonical}). However, the stationary distribution
for the system with $\phi > 0$ is not generally obtained.
Nevertheless, 
in the linear response regime out of equilibrium,
there is a useful expression
called the Zubarev-McLennan representation
\begin{equation}
  {\cal P}_{\ss}(\bv{\rho};\bar\rho, \phi)
  =  {\cal P}_{\eq}(\bv{\rho};\bar \rho)
  \exp \left[ - \frac{\phi \bra Q  \ket^{\eq}_{\bv{\rho}} +O(\phi^2) }
    { \Teff}  \right]
\label{ZM}
\end{equation}
with 
\begin{equation}
  Q=\int_0^\infty dt~ j_{N}(t),
\label{Q-def}  
\end{equation}  
where $j_N(t)$ is a fluctuating current at the $N$-th bond,
which is defined in \eqref{model:cur}.  The $N$-th bond is
the only bond on which the driving $\phi$ is imposed.
$\bra \ \ket^{\eq}_{\bv{\rho}}$ denotes the conditioned expectation for
the equilibrium path ensemble provided that the initial density profile is
given by the specified $\bv{\rho}$ as the 
argument of the stationary distribution
\cite{Zubarev,McLennan,Crooks,fh-zubarev,KN,Maes-rep}.
See Appendix \ref{sec:ZM} for the derivation of (\ref{ZM}).

For the general expression (\ref{ZM}) with (\ref{Q-def}),
we take the limit $\keff \to 0$ and consider the singular continuum
description introduced in the previous section.
From (\ref{canonical}) and (\ref{ZM}), we have
\begin{equation}
  {\cal P}_{\ss}(\bv{\rho};\bar\rho, \phi)
  = 
  \exp \left[ - \frac{1}{ \Teff}
    ( {\cal F}(\bv{\rho})+ \phi  \bra Q  \ket^{\eq}_{\bv{\rho}}
     +O(\phi^2) +{\rm const}  )
    \right]
   \delta\left( \int_0^1 dx \rho(x)-\bar \rho \right)
\label{ZM-2}
\end{equation}
with
\begin{equation}
  Q=\red{L}\int_0^\infty dt~ j(0,t).
\label{Q-def-2}  
\end{equation}  
Note that $j_N(t)$ becomes $j(0,t)=j(1,t)$ in the singular
continuum description.  The most probable density profile  is identified
as $\rho_{X_*}^\phi(x)$ that minimizes
$ -\log {\cal P}_\ss(\bv{\rho}_X^{\phi};\bar \rho,\phi)$.
This is the variational principle to determine $X_*$.
From (\ref{ZM-2}), we obtain the variational function as 
\begin{equation}
  \calV (X;\bar\rho,\phi)=\calF(\bv{\rho}_X^\phi)+
  \phi \bra Q \ket_{\bv{\rho}_X^\phi}^\eq +O(\phi^2).
\label{V-exp}
\end{equation}
We then find $X_*$ as the special value of $X$ 
that minimizes the variational function
$\calV(X;\bar \rho, \phi)$ with $(\bar\rho,\phi)$ fixed.
That is,  $X_*$ is determined as
  \begin{equation}
 \calV (X_*;\bar\rho,\phi) =\min_{X}  \calV (X;\bar\rho,\phi) .
 \label{V-exp-2}
 \end{equation}
We thus need to calculate $\bra Q \ket_{\bv{\rho}_X^\phi}^\eq$.

\subsection{Preliminaries for the calculation
  $\bra Q \ket_{\bv{\rho}_X^{\phi}}^\eq$}
\label{Cal-Q}

\begin{figure}[t]
\centering
\includegraphics[scale=0.65]{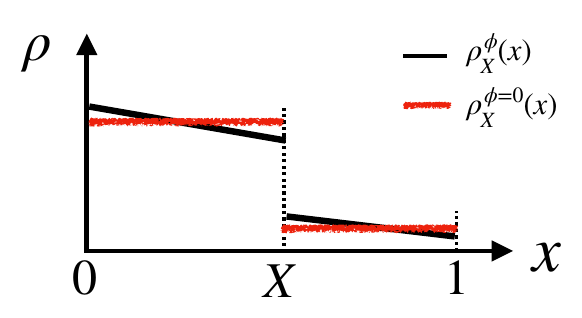}
\caption{
Schematic relationship between $\rho_X^\phi(x)$ and $\rho_X^{\phi=0}(x)$.}
\label{fig:rhoX}
\end{figure}

To calculate $\bra Q \ket_{\bv{\rho}_X^\phi}^{\rm eq}$, 
we have to know the typical time evolution of $\rho(x,t)$ starting from
$\rho_X^\phi(x)$ at $t=0$ under the equilibrium condition $\phi=0$
in the weak noise
limit $\Teff \to 0$, where $\Teff \to 0$ is taken after $t \to \infty$ is
considered. Here, from (\ref{V-exp}), we find that 
only $\phi$-independent terms of $\bra Q \ket_{\bv{\rho}_X^\phi}^\eq$
are necessary for the calculation because $\phi$-dependent terms are
absorbed into $O(\phi^2)$. We then notice expansions
\begin{equation}
\mu_X^{\phi}(x)=\mu_X^{\phi=0}(x) + O(\phi), \qquad
\rho_X^{\phi}(x)=\rho_X^{\phi=0}(x) + O(\phi). 
\label{rho-exp}
\end{equation}
See Fig. \ref{fig:rhoX} for illustration of $\rho_X^{\phi}(x)$ and
$\rho_X^{\phi=0}(x)$. See also the sentence involving (\ref{mueq})
for $\mu_X^{\phi=0}$.  
Using these relations, we find 
\begin{equation}
\bra Q \ket_{\bv{\rho}_X^\phi}^{\rm eq}= 
\bra Q \ket_{\bv{\rho}_X^{\phi=0}}^\eq +O(\phi) .
\end{equation}
That is, we study the typical time evolution of the density field starting from 
$\rho_X^{\phi=0}(x)$. Since we study the case $\keff \to 0$,
$\rho_X^{\phi=0}(x)$ represents a metastable profile that does not evolve
in time without noises. 
Nevertheless, weak noise  slowly drives the metastable profile
$\rho_X^{\phi=0}(x)$ to the equilibrium state characterized by
\eqref{canonical}.  One may conjecture that the
most probable time evolution from $\rho_X^{\phi=0}(x)$
in the weak noise limit is described by the relaxation process
    to the equilibrium profile $\rho_{X_*}^{\phi=0}(x)$
   in Fig. \ref{fig:metasta}(c) 
   from the metastable profile in Fig. \ref{fig:metasta}(b).
   However, this is not correct because of the space translational symmetry
  of equilibrium systems.  Note that metastable profiles for a given width
  $X$ of the liquid region form  a one-parameter
  family of profiles obtained by any space translation
  of $\rho_X^{\phi=0}(x)$, and
  similarly  equilibrium profiles also form  a one-parameter family. 
  Thus, stochastic dynamics in this neutral direction are equally probable,
  which are represented by  Brownian motion of the profile with the width
  of the liquid region fixed. 
  In other words, most probable  process is not uniquely determined
  in the weak noise limit. We have to
  consider a collection of trajectories that are equally probable
  and dominantly contribute to $\bra Q \ket_{\bv{\rho}_X^{\phi=0}}^\eq$.
  We call this collection {\it the highly probable path ensemble}
  to distinguish it with the most probable process.

Here, we concretely describe the highly probable path ensemble
starting from the initial condition $\rho_X^{\phi=0}(x)$.
We assume that each trajectory in the highly probable path ensemble
satisfies the following two conditions. First, the liquid or gas region
is not separated into smaller pieces of liquid or gas at any time $t$.
That is, the number of interfaces in the system is always  two.
\red{Second, slow dynamics
occurs along a continuous family of metastable states by the 
influence of weak noise, where the metastable states are characterized
by the position of two interfaces.}
We then consider the time evolution $\rho(x,t)$ as
follows. 
Let $\DL(t)$ and $\DG(t)$ denote the liquid region and gas region at
any time $t$. Recalling the relation (\ref{rhoXeq}) with (\ref{mueq}),
we express $\rho(x,t)$ in the highly probable
path ensemble as
\begin{equation}
  \rho(x,t)= \rho^{\rm L}(t) \chi(x \in \DL(t))
  + \rho^{\rm G}(t) \chi(x \in \DG(t)),
\label{assump}  
\end{equation}  
where 
$\rho^{\rm L}(t)$ and $\rho^{\rm G}(t)$
are determined from
\begin{equation}
  \mu(\rho^{\rm L}(t))=\mu(\rho^{\rm G}(t))
\end{equation}
and 
\begin{equation}
  \rhoL(t) |\DL(t)|+  \rhoG(t) |\DG(t)|=\bar \rho.
\end{equation}
It should be noted that we adopt the singular continuum description introduced
in the previous section. Using this form
of the time evolution, we estimate (\ref{Q-def}). 

Because $\rho(x,t)$ in (\ref{assump}) is described by $\DL(t)$ and
$\DG(t)$, we explicitly express them in terms of 
 the interface positions
$\Yu(t)$ and $\Yd(t)$ at time $t$, where they satisfy
\begin{equation}
  \rho(\Yu(t)+\ep,t) - \rho(\Yu(t)-\ep,t) >0,
\quad  \rho(\Yd(t)+\ep,t) - \rho(\Yd(t)-\ep,t) < 0
\end{equation}
for small positive $\ep$. See Fig. \ref{fig:psi}(a).
Then, 
the liquid region $\DL(t)$ and gas region $\DG(t)$
are expressed as 
$\DL(t)=[\Yu(t),\Yd(t)]$ and $\DG(t)=[0,1] \backslash  \DL(t)$
if $\Yu(t) < \Yd(t)$, or $\DG(t)=[\Yd(t),\Yu(t)]$
and $\DL(t)=[0,1] \backslash \DG(t)$ if $\Yu(t) > \Yd(t)$.

Now, we consider the time evolution of interface positions,
$\Yu(t)$ and $\Yd(t)$, starting from $\Yu(0)=0$ and $\Yd(0)=X$.
However, because $0 \le \Yu(t) \le 1 $ and $0 \le \Yd(t) \le 1$,
$\Yu(t)$ and $\Yd(t)$ are not continuous functions of $t$. This
would lead to a complicated calculation of the accumulated current
defined by (\ref{Q-def-2}). To describe the interface motion using
continuous functions,
we introduce generalized coordinates  $\hatYu(t) \in \mathbb{R}$
and $\hatYd(t) \in \mathbb{R}$ such that 
displacements of the interface from the initial time $0$ to
the time $t$ are given by $\hatYu(t)-\hatYu(0)$ and $\hatYd(t)-\hatYd(0)$. 
That is, $\hatYu(t)$ and $\hatYd(t)$ describe the positions
of the left and right interfaces of the liquid region
in a generalized coordinate space $\mathbb{R}$. 
The interface positions $\Yu(t)$ and $\Yd(t)$ in the space $[0,1]$
are then obtained as 
\begin{align}
\Yu(t)=\hatYu(t)-\lfloor \hatYu(t) \rfloor, \qquad
\Yd(t)=\hatYd(t)-\lfloor \hatYd(t) \rfloor,
\end{align}
where $\lfloor \ \rfloor$ represents the floor function.
The width of the liquid and gas regions are then written as
\begin{align}
&|\DL(t)|=\hatYd(t)-\hatYu(t), \quad
|\DG(t)|=1-|\DL(t)|=\hatYu(t)-\hatYd(t)+1
\label{e:DL-DG}
\end{align}
irrespective of the sign of $\Yd(t)-\Yu(t)$.  
We also introduce the center of the liquid region
in the generalized coordinate space as 
\begin{equation}
\Ys(t)=\frac{\hatYu(t)+\hatYd(t)}{2}.  
\label{Ys-def}
\end{equation}
Using the width $|\DL(t)|$ and the center $\Ys(t)$ of
the liquid region, we have
\begin{align}
\hatYd(t)=\Ys(t)+\frac{|\DL(t)|}{2},
\qquad
\hatYu(t)=\Ys(t)-\frac{|\DL(t)|}{2}.
\label{e:Yd-Yu}
\end{align}
We note that, in the weak noise limit, \red{$|\DL(t)|$ obeys
a deterministic equation describing} 
$|\DL(t)| \to X^{\rm eq}$ as $t \to \infty$,
while $\Ys(t)$  shows unbounded-free Brownian motion because of the
translation symmetry for the case $\phi=0$. This fact simplifies
the calculation of the accumulated current defined by (\ref{Q-def}).

At the end of this subsection,  we discuss the difference between
the most probable process for the case $\keff \gg 1$ and the highly
probable path ensemble for the case $\keff \ll 1$. In the former
case, $\rho_X^{\phi=0}$ evolves to an equilibrium configuration in the
deterministic system, which is in contrast with the latter case.
To obtain the accumulated current $Q$ for the
former case, we have to analyze the time-dependent solution of the
deterministic equation, which is out of the present paper.

\subsection{Expression of $j(0,t)$}
\label{cal-j}

\begin{figure}[t]
\centering
\includegraphics[scale=0.625]{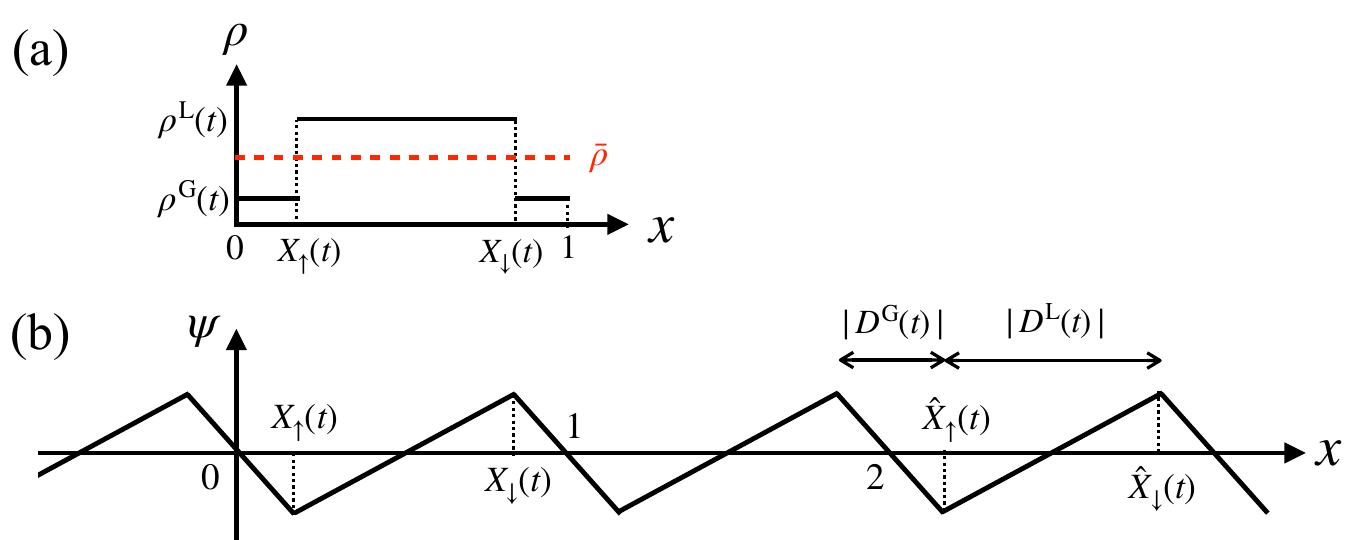}
\caption{
  (a) Density profile $\rho(x,t)$ whose interface
   positions are $\Yu(t)$ and $\Yd(t)$. 
   $\rho(x,t)=\rhoL(t)$ in $\DL(t)=[\Yu(t),\Yd(t)]$,
   while $\rho(x,t)=\rhoG(t)$ in $\DG(t)=[0,1] \backslash  \DL(t)$.
(b) Example of $\psi(x,t)$ defined in $-\infty < x < \infty$.}
\label{fig:psi}
\end{figure}

In this subsection, we calculate $j(x,t)$ in the singular continuum
description based on \eqref{continuity-hat} and \eqref{j-continuous}.
In the argument below, $\hatYu(t)$, $\hatYd(t)$, $\DL(t)$, and $\DG(t)$
are simply denoted by $\hatYu$, $\hatYd$, $\DL$, and $\DG$ if their
$t$-dependencies are clearly guessed.

For a given density profile $(\rho(x,t))_{x=0}^1$ at time $t$, 
we define
\begin{equation}
\psi(x,t)\equiv \int_0^x dx' (\rho(x',t)-\bar \rho)  
\label{e:def-psi}
\end{equation}
for $0 \le x \le 1$. 
Because $\psi(0,t)=\psi(1,t)=0$, $\psi(x,t)$ can be extended to
a periodic function in $x$. That is, we define
$\psi(x,t)\equiv \psi(x-\lfloor x \rfloor,t)$ for any $-\infty < x <\infty$.
See Fig. \ref{fig:psi}(b) for the illustration.
The time derivative of \eqref{e:def-psi} leads to
\begin{equation}
\partial_t \psi(x,t)=-j(x,t)+j(0,t) 
\label{e:psi-t}
\end{equation}
for any $x \in [0,1]$.
We here integrate \eqref{e:psi-t} over the liquid region
    and divide by $\sigmaL$. We repeat the same operation for the gas region.
Summing up the two results, 
we have a relation
\begin{equation}
\frac{1}{\sigmaL}\int_{\DL}dx \partial_t \psi
  +
\frac{1}{\sigmaG}\int_{\DG}dx \partial_t \psi
=-
\frac{1}{\sigmaL}\int_{\DL}dx j(x,t)
-
\frac{1}{\sigmaG}\int_{\DG}dx j(x,t)
+\left( \frac{|\DL|}{\sigmaL}+\frac{|\DG|}{\sigmaG} \right)
~j(0,t) .
\label{qcal-start}
\end{equation}
\red{Here, we attempt to extract slow dynamics of $j(0,t)$
by considering a short-time average of (\ref{qcal-start}).} 
Recalling the formula for $j(x,t)$ given in \eqref{j-continuous},
\red{we find that, in the weak noise limit,} 
$\mu(x,t)$ takes a constant value
in the bulk regions \red{and the space integral of the noise term 
is estimated as zero by the short-time average. Then, } the first and second terms on the right
side of \eqref{qcal-start} turn out to be zero.
We thus obtain the expression for $j(0,t)$ as
\begin{align}
j(0,t)= 
\left[
\frac{|\DL|}{\sigmaL}+\frac{|\DG|}{\sigmaG}
\right]^{-1}  
\left[
\frac{1}{\sigmaL}\int_{\DL}dx~ \partial_t \psi
  +
  \frac{1}{\sigmaG}\int_{\DG}dx~ \partial_t \psi
\right] .
\label{j-exp}
\end{align}  
We remark that $j(0,t)$ is \red{still} 
stochastic. \red{Indeed}, in the right side of \eqref{j-exp}, $\DL(t)$, $\DG(t)$,
and the space integrals of $\partial_t\psi(x,t)$ are
affected by the Brownian motion of the interfaces.


The current $j(0,t)$ formulated in \eqref{j-exp} contains the time derivative of $\psi$ in the space integrals. 
We now transform \eqref{j-exp} into a formula by letting the time derivative be outside the integral.
In the transformation procedure, we need to pay attention to the time-dependent ranges $\DL$ and $\DG$ of the integrals.  As the result,
the current $j(0,t)$ is expressed as 
\begin{equation}
j(0,t)= \Phi(t)-\Phi_0(t),
\label{j0-form}
\end{equation}  
where 
\begin{align}
\Phi(t) \equiv \frac{d}{dt}
\left\{
\left[
\frac{|\DL|}{\sigmaL}+\frac{|\DG|}{\sigmaG}
\right]^{-1}  
\left[
\frac{1}{\sigmaL}\int_{\DL}dx~  \psi
  +
  \frac{1}{\sigmaG}\int_{\DG}dx~ \psi
\right]
\right\} ,
\label{Phi-def}
\end{align}
and $\Phi_0(t)$ is determined from  \eqref{j-exp}, \eqref{j0-form},
and \eqref{Phi-def}.
Concretely, we perform the time-derivative of \eqref{Phi-def}. 
In the time derivative of a function of Brownian motion $X(t)$,
we note the chain rule $d f(X(t))/dt= f'(X(t))\circ dX/dt$, where the symbol
$\circ$ represents the Stratonovich product. We then have
\begin{equation}
\Phi_0(t)= 
\left( \frac{1}{\sigmaL}-\frac{1}{\sigmaG} \right)
\left[
\frac{|\DL(t)|}{\sigmaL}+\frac{|\DG(t)|}{\sigmaG}
\right]^{-1}  
|\DL(t)|(\rhoL(t)-\bar\rho)
\circ \frac{d}{dt} \Ys  .
\label{Phi0}
\end{equation}
See Appendix \ref{sec:Phi-Phi0} for the derivation of $\Phi_0(t)$.

Due to the translational invariance for the case $\phi=0$,
$\Ys(t)$ shows the free Brownian motion. \red{Because $|\DL(t)|$
and $\rhoL(t)$ obey deterministic equations in the weak noise limit, 
they have no correlation with $\hat Y(t)$.}
Therefore, taking the path ensemble average over noise realization,
we have
\begin{align}
\bra\Phi_0(t)\ket
&=
\left( \frac{1}{\sigmaL}-\frac{1}{\sigmaG} \right)
\bra{\left[
\frac{|\DL(t)|}{\sigmaL}+\frac{|\DG(t)|}{\sigmaG}
\right]^{-1}  
|\DL(t)|(\rhoL(t)-\bar\rho)}\ket
 \bra\frac{d}{dt} \Ys \ket \nonumber\\
&=0
\end{align}
Combining this with (\ref{j0-form}), we obtain
\begin{equation}
  \bra j(0,t)\ket^{\rm eq}_{\bv{\rho}^{\phi=0}_X}
  = \bra \Phi(t) \ket^{\rm eq}_{\bv{\rho}^{\phi=0}_X} .
\label{j-Phi}
\end{equation}

\subsection{Result of $\bra Q \ket^{\eq}_{\bv{\rho}_X^{\phi=0}}$ }
\label{result-j}

Let us define 
\begin{equation}
  q(\tau)\equiv \int_0^\tau dt~ \Phi(t).
\label{q-def}
\end{equation}  
From \red{\eqref{Q-def-2}} and (\ref{j-Phi}), we have
\begin{equation}
  \bra Q \ket^{\eq}_{\rho_X^{\phi=0}}=\red{L}
  \lim_{\tau \to \infty} \bra q(\tau) \ket^{\eq}_{\rho_X^{\phi=0}}.
\end{equation}

Substituting (\ref{Phi-def}) into (\ref{q-def})
with noting $\Yu(0)=0$ and $\Yd(0)=X$,
we obtain
\begin{align}
q(\tau) =& 
\left[
\frac{|\DL(\tau)|}{\sigmaL}+\frac{|\DG(\tau)|}{\sigmaG}
\right]^{-1}  
\left[
\frac{1}{\sigmaL}\int_{\DL(\tau)} dx~  \psi
  +
  \frac{1}{\sigmaG}\int_{\DG(\tau)} dx~ \psi
\right]   \nonumber \\
&\quad -
\left[
\frac{X}{\sigmaL}+\frac{1-X}{\sigmaG}
\right]^{-1}  
\left[
\frac{1}{\sigmaL}\int_0^X dx~  \psi
+
\frac{1}{\sigmaG}\int_X^{1} dx~ \psi
\right]   .
\label{e:q-1}
\end{align}
From the piece-wise linear nature of $\psi(x,t)$, the integrals are calculated by the trapezoidal rule as
\begin{align}
&  \int_{\DL}dx~ \psi(x,t)
= \frac{1}{2} |\DL| \left(\psi(\hatYu(t),t)+\psi(\hatYd(t),t)\right),
\label{int-psi-L} \\
&  \int_{\DG}dx~ \psi(x,t)
= \frac{1}{2} |\DG| \left(\psi(\hatYu(t),t)+\psi(\hatYd(t),t)\right),
\label{int-psi-G}
\end{align}
where $\psi(\Yu(t)+1)=\psi(\Yu(t))$ is applied.
Using these relations, we obtain
\begin{align}
\frac{1}{\sigmaL}\int_{\DL}dx~  \psi
  +
 \frac{1}{\sigmaG}\int_{\DG}dx~  \psi
 =
 \frac{1}{2}
\left(  \frac{|\DL|}{\sigmaL}+\frac{|\DG|}{\sigmaG} \right)
\left(\psi(\hatYu(t),t)+\psi(\hatYd(t),t)\right) .
\label{int-psi}
\end{align}
Substituting this into \eqref{e:q-1}, we have
\begin{align}
q(\tau)=&
\frac{1}{2}\left(\psi(\hatYu(\tau),\tau)+\psi(\hatYd(\tau),\tau)\right)
-\frac{1}{2} \left(\psi(0,0)+\psi(X,0)\right).
\end{align}
Here, 
\begin{align}
\psi(0,0)=0,\qquad \psi(X,0)=(\rhoLX-\bar\rho)X,
\end{align}
and
\begin{align}
\psi(\hatYu(\tau),\tau)=\psi(\Yu(\tau),\tau),\qquad 
\psi(\hatYd(\tau),\tau)=\psi(\Yd(\tau),\tau).
\end{align}

We thus obtain
\begin{equation}
\lim_{\tau\to\infty}\bra q(\tau)\ket^{\eq}_{\bv{\rho}_X^{\phi=0} \to *}=
 \frac{1}{2} \lim_{\tau \to \infty}\left(\bra{\psi(\Yu(\tau),\tau)}\ket +\bra{\psi(\Yd(\tau),\tau)}\ket \right)
-\frac{1}{2} (\rhoLX-\bar \rho) X.
\label{Q-result0}
\end{equation}
Note that the positions of the interfaces, $\Yu$ and $\Yd$,
are uniformly distributed in the interval $x\in[0,1]$
as $\tau\to\infty$.
We thus conclude that
\begin{equation}
\bra Q \ket^{\eq}_{\bv{\rho}_X^{\phi=0}}
=\red{L} \left[ C
-\frac{1}{2} (\rhoLX-\bar \rho) X \right],
\label{Q-result}
\end{equation}
where $C$ is a constant independent of $X$.
Using the relation
\begin{align}
\bar\rho=X\rhoLX+(1-X)\rhoGX,
\label{e:bar-rho}
\end{align}
which comes from (\ref{barrho-def}),
we can also express (\ref{Q-result}) as
\begin{equation}
\bra Q \ket^{{\eq}}_{\bv{\rho}_X^{\phi=0}}
=\red{L} \left[C
-\frac{1}{2} (\rhoLX-\rhoGX) X(1-X) \right].
\label{Q-result-symmetric}
\end{equation}

Substituting \eqref{Q-result-symmetric} into \eqref{V-exp}, we have
\begin{equation}
  \calV(X;\bar\rho,\phi)=\red{L}\left[\int_0^1 dx~ f(\rho_X^\phi(x))
 -\frac{\phi}{2}(\rhoLX-\rhoGX)X(1-X)+ \phi C \right].
 \label{V}
\end{equation}
We note that $\rho_X^\phi(x)$ is uniquely determined
for given $(X,\phi,\bar\rho)$ by $\mu(\rho_X^\phi(x))=\mu_X^\phi(x)$
with \eqref{muLc} and \eqref{muGc} and that
$\rhoLX$ and $\rhoGX$ are functions of $(X,\bar\rho)$.
Thus, $ \calV(X;\bar\rho,\phi)$ is the variational function
for determining the most probable interface position $X_*$
for a given $(\bar\rho,\phi)$. The most probable density
profile is expressed as $\rho_{X_*}^\phi(x)$.

\section{Variational equation}
\label{Val}

In this section, using the variational function \eqref{V},
we determine the most probable value of $X$.
This is regarded as an extension of the argument for determining the
equilibrium profile in the paragraph containing  \eqref{var-FX-eq}.
Thus, similarly to \eqref{var-FX-eq}, we analyze 
the variational equation
\begin{equation}
\left. \der{\calV(X;\bar \rho,\phi)}{X} \right|_{X=X_*}=0
\end{equation}  
with \eqref{V}. The difference from \eqref{var-FX-eq}
is the $O(\phi)$ terms in \eqref{V}. 
As shown in Fig. \ref{fig:profile},
one of the interfaces of $\rho_X^\phi(x)$ is located  at $x=0$
for any $X$, both the liquid and gas density profiles
are sloped, and the chemical potential profile $\mu_X^\phi(x)$
is piece-wise linear. Then, as did for the equilibrium system,
we first derive the chemical potential at the interface 
$\mu_{X_*}^\inter$ defined by 
\eqref{mu-int-def} instead of directly calculating $X_*$.
Once we have $\mu_{X_*}^\inter$,
we obtain the most probable profiles $\mu_{X_*}^\phi(x)$ and
$\rho_{X_*}^\phi(x)$, and the value of $X_*$. The determination
of $\mu_{X_*}^\inter$ is also physically important because 
if $\mu_{X_*}^\inter \neq \muc$,
metastable states at equilibrium stably appear in the
non-equilibrium phase coexistence.

\subsection{Preliminaries for the calculation}
\label{prel-var}

In the argument below, we ignore the contribution of $O(\phi^2)$.
We first define
\begin{equation}
\barrhoLX \equiv \frac{1}{X}\int_0^X dx \rho^\phi_X(x), \quad  
\barrhoGX \equiv \frac{1}{1-X}\int_X^1 dx \rho^\phi_X(x),
\end{equation}
and
\begin{equation}
\barmuLX \equiv \frac{1}{X}\int_0^X dx \mu^\phi_X(x), \quad  
\barmuGX \equiv \frac{1}{1-X}\int_X^1 dx \mu^\phi_X(x).
\end{equation}
Because the density profile $\rho_X^\phi(x)$ and the chemical potential
 profile $\mu_X^\phi(x)$ are linear in the respective regions,
$[0,X]$ and $[X,1]$, 
we obtain
\begin{equation}
f\left( \barrhoLX \right)=\frac{1}{X}\int_0^X dx f(\rho^\phi_X(x)),\qquad
f\left( \barrhoGX \right)=\frac{1}{1-X}\int_X^1 dx f(\rho^\phi_X(x)),
\end{equation}
and
\begin{equation}
\barmuLX=\mu(\barrhoLX), \qquad \barmuGX=\mu(\barrhoGX).
\label{91}
\end{equation}
The pressures in the liquid and gas regions are characterized
  by $\pLX \equiv p(\barrhoLX)$ and $\pGX \equiv p(\barrhoGX)$, which are
  expressed by 
\begin{align}
\pLX=\muLX\barrhoLX-f(\barrhoLX), \qquad
\pGX=\muGX\barrhoGX-f(\barrhoGX).
\label{e:P-def}
\end{align}

The first term on the right side of (\ref{V}) is written as
\begin{equation}
\int_0^1 dx~ f(\rho_X^\phi(x))=X f\left( \barrhoLX \right)
+(1-X)  f\left( \barrhoGX \right)+O(\phi^2),
\label{mu-rho-f}
\end{equation}
and the variational function \eqref{V-exp} as
\begin{align}
{\calV}(X;\bar \rho, \phi)=\red{L}\left[
X f\left( \barrhoLX \right)+(1-X)  f\left(\barrhoGX \right)
-\frac{\phi}{2}\left(\rhoLX-\rhoGX\right)X(1-X)+\phi C\right].
\label{var-ft-0}
\end{align}
Since $\barrhoLX=\rhoLX+O(\phi)$ and 
$\barrhoGX=\rhoGX+O(\phi)$, we can rewrite (\ref{var-ft-0}) as
\begin{align}
{\calV}(X;\bar \rho, \phi)=\red{L}\left[
X f\left( \barrhoLX \right)+(1-X)  f\left(\barrhoGX \right)
-\frac{\phi}{2}\left(\barrhoLX-\barrhoGX\right)X(1-X)+\phi C\right].
\label{var-ft}
\end{align}
We emphasize that \eqref{var-ft} is explicitly expressed  as a function of $X$.

Noting that the chemical potential profile is piece-wise linear as shown
in  Fig. \ref{fig:profile},
 we estimate
\begin{align}
\barmuLX= \frac{\mu_X^\phi(0)+  \mu_X^\inter}{2}, \qquad 
\barmuGX=  \frac{\mu_X^\phi(1)+  \mu_X^\inter}{2}.
\label{star}
\end{align}
From \eqref{star},  $\phi=\mu_X^\phi(1)-\mu_X^\phi(0)$ is rewritten as
\begin{equation}
\barmuLX - \barmuGX=\frac{\phi}{2}.
\label{e:phi-LG}
\end{equation}
Furthermore, using \eqref{muLc} and \eqref{muGc},
we can express $\barmuLX$ and $\barmuGX$ in terms
of $\mu_X^\inter$ as
\begin{align}
\barmuLX  =\mu_X^\inter +\frac{JL}{\sigmaL}\frac{X}{2}, \quad
\barmuGX  =\mu_X^\inter -\frac{JL}{\sigmaG}\frac{1-X}{2}.
\label{e:muLG-J}
\end{align}

\subsection{Steady state}

We consider the variational equation 
\begin{equation}
\der{\calV(X;\bar \rho,\phi)}{X} =0.
\label{v-eq}
\end{equation} 
Substituting \eqref{var-ft} into \eqref{v-eq}, we have
\begin{align}
 & f(\barrhoLX)-f(\barrhoGX)-\frac{\phi}{2}(\barrhoLX-\barrhoGX)(1-2X) \nonumber \\
+& \barmuLX \frac{d\barrhoLX}{dX}X+\barmuGX \frac{d\barrhoGX}{dX}(1-X)
-\frac{\phi}{2}\left(\frac{d\barrhoLX}{dX}-\frac{d\barrhoGX}{dX}\right)X(1-X)
\nonumber \\
=& 0. 
\label{start}
\end{align}
Using \eqref{e:phi-LG}, we rewrite  the second line as
\begin{align}
[X\barmuLX+(1-X)\barmuGX]\left[X\frac{d\barrhoLX}{dX}+(1-X)\frac{d\barrhoGX}{dX}\right].
\label{e:veq-1}
\end{align}
Here, taking the derivative of $\bar \rho=X \barrhoLX+(1-X)\barrhoGX$ in $X$,
we obtain 
\begin{align}
X\frac{d\barrhoLX}{dX}+(1-X)\frac{d\barrhoGX}{d\red{X}}=-(\barrhoLX-\barrhoGX).
\end{align}
We substitute this into \eqref{e:veq-1} and combine it with
(\ref{start}).  Then, the variational equation \eqref{start} becomes 
\begin{align}
f(\barrhoLX)-f(\barrhoGX)-(\barrhoLX-\barrhoGX)[(1-X)\barmuLX+X\barmuGX]=0.
\label{e:veq-2}
\end{align}
The solution of \eqref{e:veq-2} provides the most probable value
$X_*$ of the interface position. Therefore, we express
\begin{align}
f(\barrhoLXs)-f(\barrhoGXs)-(\barrhoLXs-\barrhoGXs)[(1-X_*)\barmuLXs+X\barmuGXs]=0.
\label{e:veq-22}
\end{align}

Now, we rewrite  (\ref{e:veq-22}) as a different form using the chemical
potential at the interface position $X_*$.
We subtract the equilibrium version of \eqref{e:veq-22},
i.e., $f(\rhoLc)-f(\rhoGc)-(\rhoLc-\rhoGc)\muc=0$, from \eqref{e:veq-22}.
Noting $f(\barrhoLXs)-f(\rhoLc)=(\barrhoLXs-\rhoLc)\muc+O(\phi^2)$,
we obtain
\begin{align}
(\barrhoLXs-\barrhoGXs)\left\{\muc-[(1-X_*)\barmuLXs+X_*\barmuGXs]\right\}=0.
\label{e:v-eq-final}
\end{align}
Substituting \eqref{e:muLG-J} into this equation and 
using $\barrhoLXs\neq\barrhoGXs$, we have
\begin{equation}
\mu_{X_*}^\inter=
\muc-\frac{ JL X_*(1-X_*)}{2} \left( \frac{1}{\sigmaL}
- \frac{1}{\sigmaG}\right).
\end{equation}
Because $X_*=\Xeq+O(\phi)$, we can rewrite it as
\begin{equation}
\mu_{X_*}^\inter=
\muc-\frac{ JL \Xeq(1-\Xeq)}{2}
\left( \frac{1}{\sigmaL} - \frac{1}{\sigmaG}\right).
\label{100}
\end{equation}
Furthermore, combining \eqref{J-phi} with
\eqref{100}, we finally obtain 
\begin{equation}
\mu_{X_*}^\inter =
\muc+\frac{\phi}{2} \frac{(\sigmaL-\sigmaG)\Xeq(1-\Xeq)}
{\sigmaG \Xeq+\sigmaL (1-\Xeq)}.
\label{muI-final}
\end{equation}
Recalling that $\Xeq$ is uniquely determined by $\bar \rho$
from \eqref{Xeq-det}, we conclude that $\mu_{X_*}^\inter$ is
expressed in terms of the system parameters.
Thus, the chemical potential at the interface deviates
from $\muc$ linearly with the voltage $\phi$. This means
that metastable states at equilibrium stably appear
around the interface. The relation \eqref{muI-final}
is the most important achievement of our theory.

Next, from \eqref{e:phi-LG} and \eqref{e:v-eq-final}, we obtain
\begin{align}
\barmuLXs  =\muc +\Xeq\frac{\phi}{2}, \quad
\barmuGXs  =\muc -(1-\Xeq)\frac{\phi}{2},
\label{e:muLG-c}
\end{align}
which also gives $\mu_{X_*}^\phi(0)$ and $\mu_{X_*}^\phi(1)$
using \eqref{star}.
The result yields $\mu_{X_*}^\phi(x)$. Using
$\mu_{X_*}^\phi(x)=\mu(\rho_{X_*}^\phi(x))$, we have $\rho_{X_*}^\phi(x)$.
Finally, from $ X_* \barrhoLXs+(1-X_*) \barrhoGXs =\bar \rho$, we can
express $X_*-\Xeq$ in terms of system parameters. In this manner, all
thermodynamic quantities are determined by the variational principle.
As one example, 
we discuss the pressure in the steady state.
Using \eqref{e:P-def}, we can express \eqref{e:veq-22} as
\begin{align}
\pLXs-\pGXs=\frac{\phi\bar\rho}{2}.
\label{e:p-relation}
\end{align}
Recalling that the local pressure is given by $p(x)=p(\rho(x))$,
we define the pressures at the left and right sides of the
interface as 
\begin{align}
p_{-}\equiv \lim_{x\to X_*^-}p(\rho(x)),\qquad
p_{+}\equiv \lim_{x\to X_*^+}p(\rho(x)).
\end{align}
Then, using
\begin{equation}
  \rho^{\rm L/G}_{X_*} (\mu^{\rm L/G}_{X_*}-\mu^{\rm I}_{X_*})
  =p^{\rm L/G}_{X_*}- p_{-/+} ,
\end{equation}
we can derive
\begin{align}
p_-=p_{\rm c}+(\mu_{X_*}^\inter-\muc)\rhoLc,
\qquad
p_+=p_{\rm c}+(\mu_{X_*}^\inter-\muc)\rhoGc,
\label{pdiff}
\end{align}
where $p_{\rm c}$ is the equilibrium coexistence pressure.
This result indicates that the pressure is not continuous
at the interface. It should be noted that this discontinuity
occurs only in the singular continuum description for $\keff \ll 1$ but
never occur in the continuum description for $\keff \gg 1$ as shown
in Appendix \ref{app:con}.

\section{Global thermodynamics}
\label{Global}

\red{
The extension of the variational function
from ${\cal F}_\eq(X;  \bar \rho) $ to ${\cal F}_\ss(X;  \bar \rho, \phi)$ is closely related to the extension of thermodynamic functions.
Without analyzing specific
stochastic models, one can construct such an extended thermodynamic
framework relying on the consistency, uniqueness, and predictability.
This phenomenological argument, which is called {\it global thermodynamics},
was developed for heat conduction systems exhibiting phase coexistence
\cite{Global-PRL,Global-JSP,Global-PRR}. Furthermore, global
thermodynamics was applied to the order-disorder transition in heat
conduction, and the prediction by global thermodynamics
was confirmed by numerical simulations \cite{KNS}. 
Similarly, in the present setup, 
we can determine the variational function \eqref{var-ft}
following the method in Refs. 
\cite{Global-PRR,Global-gravity}. 
}

We set $ M \equiv  \bar \rho L$, where $MA$ represents
the total number of particles in the tube.
In equilibrium thermodynamics, 
the free energy function $F_\eq(L,M)$ is determined as
\begin{equation}
  F_\eq(L,M)={\cal F}_\eq(X_*;\bar \rho)
  \label{eq:free}
\end{equation}
using the variational function ${\cal F}_\eq(X;\bar \rho)$
given by \eqref{eq-var}. 
We then have the fundamental relation of thermodynamics
\begin{equation}
  dF_\eq= -\pc dL+\muc dM,
\label{eq:thermo}
\end{equation}
where $\pc$ and $\muc$ are the equilibrium values of pressure
and chemical potential \red{in phase coexistence states.
Note that $\pc$ and $\muc$ are constants in $(L,M)$.
Extending  the relations \eqref{eq:free} and \eqref{eq:thermo} 
to non-equilibrium systems, we attempt to determine the thermodynamic
function $F_\ss(L,M,\phi)$ and the variational function 
${\cal F}_\ss(X;\bar \rho,\phi)$. } 

\red{
 \subsection{Thermodynamic function}
\label{ap:thermo}
}

\red{
For the system under consideration, we have the equilibrium free 
energy $F_{\rm eq}(L,M)$. In the phase coexistence state, 
it takes the simple form
\begin{equation}
F_{\rm eq}(L,M)=-\pc L+\muc M .  
\label{feq-0}
\end{equation}
To derive $F_\ss(L,M,\phi)$,
we first assume 
\begin{align}
  dF_\ss = -\bar p dL +\bar \mu d M -\Psi d\phi.
  \label{fund-rel}
  \end{align}
without an explicit form of $\Psi$, 
where $\bar p$ and $\bar \mu$ are the average
pressure and chemical potential defined by 
\begin{align}
  \bar p    = X_* \pLXs+ (1-X_*) \pGXs, \qquad
  \bar \mu = X_* \barmuLXs+(1-X_*) \barmuGXs.
  \label{e:tilde-mu-gl}
\end{align}
Note that $\bar p$ and $\bar \mu$ are 
functions of $(\bar \rho, \phi)$ and that 
 the steady state value $X_*$ of the interface position
is also given as a function of $(\bar \rho, \phi)$. 
$\Psi$ is a conjugate
variable of the battery voltage $\phi$. From the reflection
symmetry, $\Psi$ is an odd function of $\phi$ and thus $\Psi=0$ 
for $\phi=0$. On the other hand, from the singular nature of the
phase coexistence, we assume that $\Psi \not = 0$ in the limit 
$\phi \to 0+$. For simplicity, we consider only the case $\phi>0$
as in the previous sections and we can set $\Psi=\Psi(L,M)$ by considering 
the limit $\phi \to 0+$  to study the linear response regime.
Furthermore, from the extensivity in the equilibrium state, 
we have
\begin{align}
\Psi(L,M)=M\psi(v)  
\end{align}
with $v=L/M=1/\bar \rho$. 
} 

\red{
Next,  we set
\begin{equation}
F_{\rm ss}(L,M,\phi)=F_{\rm eq}(L,M) - M\psi(v) \phi 
\label{Fss-start}
\end{equation}
in the linear response regime. By substituting this
expression into \eqref{fund-rel}, we obtain
\begin{align}
  \bar p=\pc+ \psi'(v) \phi,  \qquad 
  \bar \mu=\muc+ (v \psi'(v)-\psi(v)) \phi.  
  \label{eq-psi}
\end{align}
We solve these differential equations of $\psi(v)$.
Eliminating $\psi'(v)$ from these two
equations, we obtain a necessary condition for $\psi(v)$
as 
\begin{align}
\psi(v) \phi = v (\bar p -\pc) -(\bar \mu-\muc) . 
\label{eq-psi-2}
\end{align}
}

\red{
Here, we simplify the right-hand side of (\ref{eq-psi-2})
using the expressions given in  \eqref{e:tilde-mu-gl}. We first
note the relations
\begin{align}
  \pLXs-\pc=\barrhoLXs (\barmuLXs -\muc) , \qquad
  \pGXs-\pc=\barrhoGXs (\barmuGXs -\muc) .
\end{align}
Using the relations,
we express (\ref{eq-psi-2}) as
\begin{align}
  \psi(v)\phi=(v \barrhoLXs-1)\Xs\barmuLXs+
  (v \barrhoGXs-1)(1-\Xs)\barmuGXs  .
  \label{eq-psi-2-R}
\end{align}
We next notice that the relation \eqref{e:bar-rho} yields useful expressions
\begin{align}
  (v \barrhoLXs-1)\Xs + (v \barrhoGXs-1)(1-\Xs)\barmuGXs=0, 
\label{u-1}
\end{align}
and
\begin{align}
  \barrhoLXs-\bar \rho=(1-\Xs)(\barrhoLXs-\barrhoGXs).
\label{u-2}
\end{align}
Using \eqref{e:phi-LG} and \eqref{u-1}, we rewrite \eqref{eq-psi-2-R} as
\begin{align}
  \psi(v)\phi=(v \barrhoLXs-1)\Xs \frac{\phi}{2}.
  \label{eq-psi-2-R2}
\end{align}
By substituting \eqref{u-2} into \eqref{eq-psi-2-R2}, we obtain
\begin{align}
\phi \Psi(L,M)= \phi M \psi(v)= 
  \frac{\phi}{2} L(\barrhoLXs-\barrhoGXs) X_*(1-X_*)  .
  \label{eq-psi-2-R3}
\end{align}
Thus, if a solution
exists for \eqref{eq-psi}, this should be 
\eqref{eq-psi-2-R3}. }

\red{
Conversely,  suppose that \eqref{eq-psi-2-R3} holds. 
Noting the relation
\begin{align}
F_{\rm eq}(L,M)=L X_*f(\barrhoLXs) +L (1-X_*)f(\barrhoGXs)
+O(\phi^2),
\end{align}
we find that
$F_{\rm ss}$ assumed in \eqref{Fss-start} 
is expressed as
\begin{align}
 F_{\ss}(L, M, \phi)&=L X_*f(\barrhoLXs) +L (1-X_*)f(\barrhoGXs)
 -\frac{\phi}{2} L(\barrhoLXs-\barrhoGXs) X_*(1-X_*)  .
 \label{GT-free-s}
 \end{align}
 }
 Then, using \eqref{e:P-def} and \eqref{e:phi-LG},
 we rewrite \eqref{GT-free-s} as
 \begin{align}
 F_{\ss}(L, M, \phi)&=- L X_* \pLXs- L(1-X_*)\pGXs+  
 (X_* \barmuLXs+(1-X) \barmuGXs) M .
 \end{align}
Furthermore, using \eqref{e:tilde-mu-gl},
we further rewrite $F_\ss$ as a suggestive form
 \begin{align}
 F_{\ss}(L, M, \phi)&=-  \bar p L +  \bar \mu M.
 \label{FGT-LM}
 \end{align}
 Now, taking the derivative of $F_\ss$ in $L$, we have
 \begin{equation}
   \pder{F_\ss(L,M,\phi)}{L}= -\bar p
      - \pder{\bar p (L,M,\phi)}{L} L 
      + \pder{\bar \mu (L,M, \phi)}{L} M     .
 \label{FGT-L}     
 \end{equation}
 Here, using \eqref{91}, \eqref{e:P-def}, \eqref{e:phi-LG}, and
 \eqref{e:p-relation}, we can confirm 
 \begin{equation}
 \pder{\bar p (L,M,\phi)}{L} 
      = \bar \rho \pder{\bar \mu (L,M, \phi)}{L}. 
 \label{key}  
 \end{equation}
 Substituting this result into \eqref{FGT-L}, we obtain
 \begin{equation}
   \pder{F_\ss(L,M,\phi)}{L}= -\bar p.
 \label{f-1}
 \end{equation}
 By repeating the similar calculation, we also have
 \begin{equation}
   \pder{F_\ss(L,M,\phi)}{M}= \bar \mu.
 \label{f-2}
 \end{equation}
 Finally, \red{ \eqref{eq-psi-2-R3} leads to}
  \begin{equation}
   \Psi = - \pder{F_\ss(L,M,\phi)}{\phi}.
 \label{f-3}
 \end{equation}
 These three relations \eqref{f-1}, \eqref{f-2}, and \eqref{f-3} are
 summarized as an extended form of the fundamental relation of
 thermodynamics \eqref{fund-rel}. \red{
This means that $\psi(v)$ given by \eqref{eq-psi-2-R3}
satisfies \eqref{eq-psi}. 
We then conclude that \eqref{GT-free-s}
is the free energy extended to the non-equilibrium steady
state. }

\red{
\subsection{Variational function}
\label{ap:var}
}

\red{
We formulate a variational principle for 
determining the steady state. In the phase coexistence state,
unconstrained thermodynamic variables are the length of the liquid region 
$L X$ and the particle number of the liquid $\ML\equiv L \rho^L$ 
per unit area.  Let ${\cal F}_{\ss} (LX, \ML; L, M, \phi )$ be
the variational function for $(LX, \ML)$ with $(L,M,\phi)$ fixed.
Following the standard method of thermodynamics, we assume 
\begin{align}
F_{\ss}(L,M,\phi)=\min_{X, \ML}{\cal F}_{\ss} (LX, \ML; L, M, \phi ).
\label{v-ss}
\end{align}
With \eqref{GT-free-s}, \eqref{v-ss} naturally leads to
\begin{align}
  {\cal F}_{\ss} (LX, \ML; L, M, \phi )
= F_{\rm eq}(LX,\ML)+F_{\rm eq}(L(1-X),\ML)
 -\frac{\phi}{2} (\ML-MX)  ,
\label{E14}
\end{align}
where we have used \eqref{eq-psi-2-R2}.
The variational equations are then expressed as
\begin{equation}
  \pder{ {\cal F}_\ss(LX, \ML; L,M, \phi)}{X}=0 ,
  \label{ap:ve-1}
  \end{equation}
  and
\begin{equation}
  \pder{ {\cal F}_\ss(LX, \ML; L,M,\phi)}{\ML}=0.
  \label{ap:ve-2}  
\end{equation}
The solution $(X_*, \ML_*)$ of the equations is the 
steady state value. 
}

\red{
Now, we construct  a variational function for determining $X_*$
from the variational function \eqref{E14} for determining 
$X_*$ and $\ML_*$. For a given $X$, we write the 
solution of \eqref{ap:ve-2} as $\MLX$. We can then confirm  
\begin{equation}
  \left. \der{ {\cal F}_\ss(LX, \MLX; L,M, \phi)}{X} \right\vert_{X=X_*}=0 
  \label{ap:ve-11}
  \end{equation}
using \eqref{ap:ve-1} and \eqref{ap:ve-2}. The equation
\eqref{ap:ve-11} means that $ {\cal F}_{\ss} (LX, \MLX; L, M, \phi )$
is the variational function for determining $X_*$.  
We note that $\MLX= L \barrhoLX$ because  \eqref{ap:ve-2} 
leads to \eqref{e:phi-LG}.  Thus, the variational function 
$ {\cal F}_{\ss} (LX, \MLX; L, M, \phi )$
takes the same form as \eqref{V} expect for the constant term. 
Namely, the steady state determined by the phenomenological argument is equivalent to that calculated in the previous section.
Therefore, all the
prediction made by global thermodynamics for the present setup are the
same as the theoretical  result for the stochastic
model we study. In our research history,  the
variational function \eqref{var-ft} was first derived using
global thermodynamics, and after that it was re-derived by  analyzing
the stochastic model. }

\section{Numerical simulation}
\label{ns}

In this section, we perform numerical simulations of the discrete
model and compare numerical results with the theoretical predictions
presented in the previous sections. More explicitly, 
the time evolution of $(\rho_i)_{i=1}^N$ is defined by \eqref{model:con}
accompanied with the current $(j_i)_{i=1}^N$ defined by \eqref{model:cur}.
To obtain $(j_i)_{i=1}^N$ using \eqref{model:cur}, we determine
$(\tilde\mu_i)_{i=1}^N$ by \eqref{tilde-mu} with 
\begin{align}
\mu(\rho_i)=(\rho_i-0.5)(\rho_i-1.5)(\rho_i-2.5)
\end{align}
from  \eqref{free}.  We adopt a simple form of the conductivity
$\sigma(\rho)=\rho$, where we have introduced a dimensionless time
in this expression.
For this specific model, we have $\rhoLc=2.5$, $\rhoGc=0.5$, and $\muc=0$ 
from \eqref{n-con}, and  we thus obtain $\sigmaL=2.5$ and $\sigmaG=0.5$ from
\eqref{sigmaLG}. 

Recalling that the independent parameters to be specified for numerical
determination of the steady state are $(\keff, \Teff, \phi, \bar \rho, N)$
as discussed around \eqref{five}, we study the $\keff$ dependence of the
steady state with fixing the other parameter values as
\begin{equation}
 (\Teff, \phi, \bar \rho, N)=(0.002, 0.05, 1.5, 64).
\end{equation}
Since $A$ and $\Lambda$ are contained in the renormalized quantities
$\keff$ and $\Teff$, we do not need to specify the values of $A$ and
$\Lambda$, while the total volume and the total number of particles
are given by $LA$ and $\bar\rho LA$ with $L=\Lambda N$.  To numerically
solve \eqref{model:con} and \eqref{model:cur}, we adopt the Heun method
with a time step $dt=0.01\Lambda^2$, where it should be noted that
the time step $dt$ is always coupled with $\Lambda^2$ for the
time-discretized form of \eqref{model:con} and \eqref{model:cur}.

\begin{figure}[t]
\centering
\includegraphics[scale=0.6]{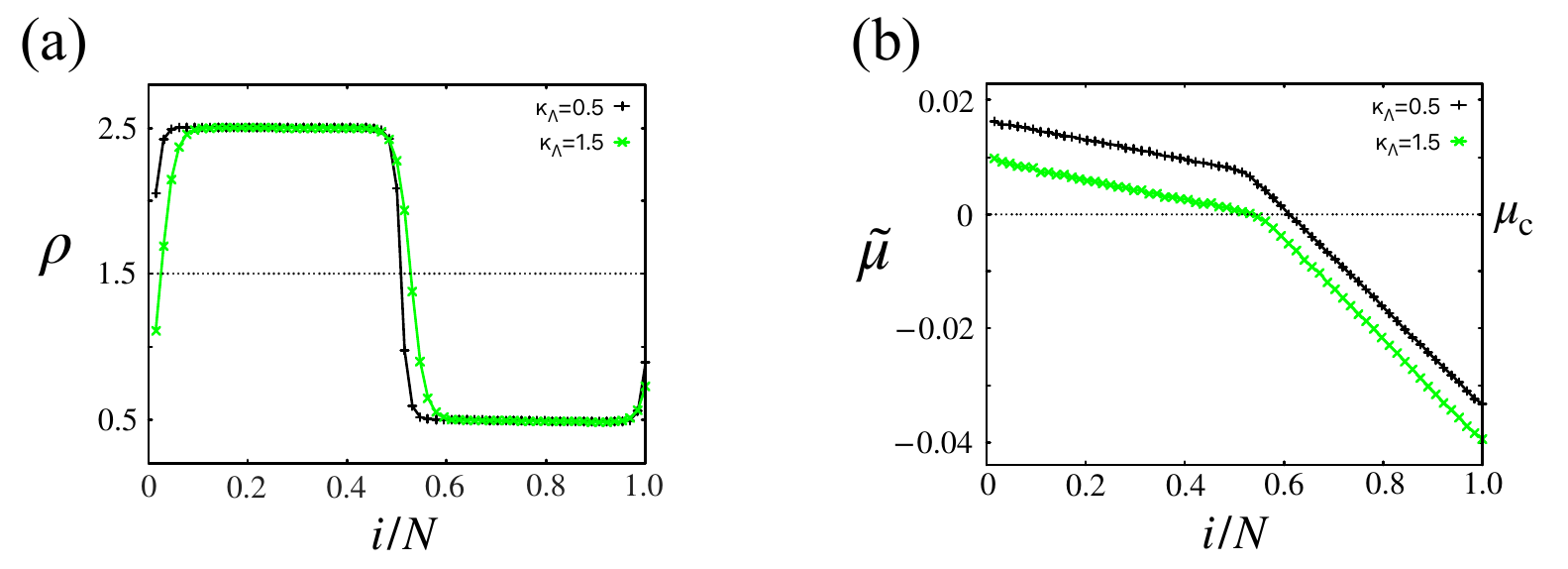}
\caption{(a) Density profile $\rho_i$ for $i/N$. (b) Chemical potential
profile $\tilde \mu_i$ for $i/N$. $\keff=0.5$ and $\keff=1.5$.
The dotted line in (b) represents the equilibrium profile of chemical
potential $\tilde \mu_i= \muc=0$.
}
\label{fig:profile-2}
\end{figure}

In Fig. \ref{fig:profile-2}, we show the steady state
for $\keff=0.5$ and $\keff=1.5$, where the density profile
$\rho_i$ and the chemical potential profile $\tilde \mu_i$ are plotted
for $i/N$. We remark that the system reaches the steady state without 
dependence on initial conditions, $(\rho_i)_{i=1}^N$ at $t=0$.
Note that $\tilde\mu_i=\muc=0$ for all $i$ for the equilibrium system
with $\phi=0$, which is shown as the dotted
line in Fig. \ref{fig:profile-2}(b).
It is observed that $\tilde \mu$ near the interface 
is close to $\muc=0$ for the case $\keff=1.5$, while it clearly deviates
from $\muc=0$ for the case $\keff=0.5$. That is, the metastable gas
stably appears at the left-side of the interface for the case $\keff=0.5$.

\begin{figure}[t]
\centering
\includegraphics[scale=0.6]{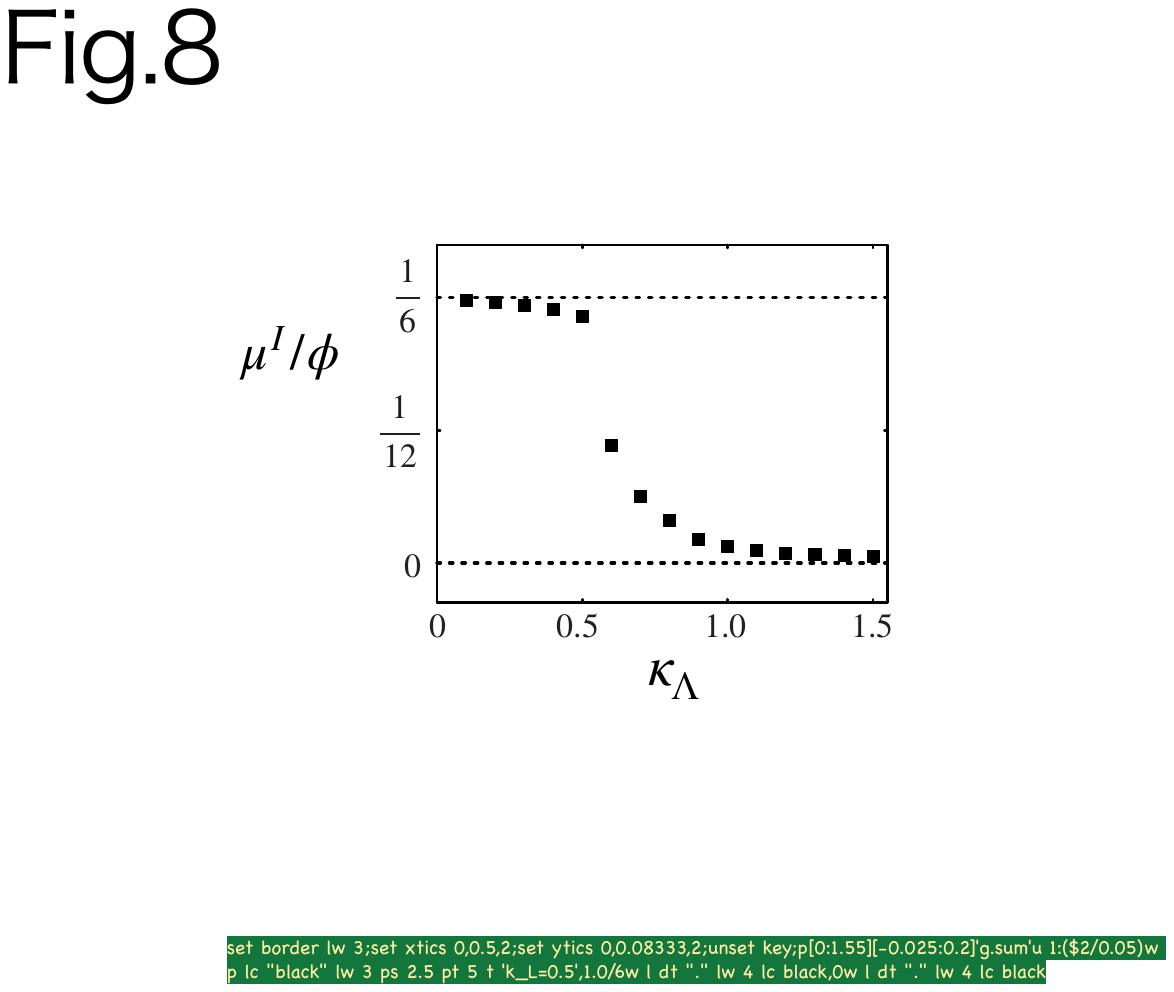}
\caption{$\keff$ dependence of $\mu^I$.
Square symbols show the numerical results for $N=64$.
The two dotted lines represent the theoretical predictions
$\mu^I =\phi/6$ for $\keff \ll 1$ and $\mu^I =0$ for $\keff \gg 1$.}
\label{fig:muI}
\end{figure}

Here, we determine the chemical potential at the interface,
which is denoted by $\mu^I$, more quantitatively from the numerical
data $(\rho_i)_{i=1}^N$ and $(\tilde \mu_i)_{i=1}^N$. In principle, we first
determine the interface position $X^I$ from the data $(\rho_i)_{i=1}^N$,
and then read the value of the chemical potential at the interface
position from the data $(\tilde \mu_i)_{i=1}^N$. In practice, we use
the linear interpolation of the data sets to systematically
estimate $\mu^I$ for several parameters. 
That is, we define a piece-wise linear function 
$\rho(x)$  for $0 \le x \le 1$ by connecting two consecutive points
$(i/N, \rho_i)$ and $((i+1)/N, \rho_{i+1})$  for $0 \le i \le N$
in the graph of $(i/N, \rho_i)_{i=1}^N$. We then define the interface
position $X^I$ as $\rho(X^I)=1.5$, where $ (\rhoGc+\rhoLc)/2=1.5$.
Similarly, we define $\tilde \mu(x)$ from $(\tilde \mu_i)_{i=1}^N$.
Using this $X^I$ and $\mu(x)$, we obtain $\mu^I =\tilde \mu(X^I)$.
More explicitly, 
$\mu^I$ is determined as follows. First, we find $i_*$ satisfying
$\rho_{i_*-1} > 1.5$ and $\rho_{i_*} <1.5$. From the construction of
$\rho(x)$, we obtain 
\begin{equation}
 X^I= \frac{1.5-\rho_{i_*}}{\rho_{i_*-1}-\rho_{i_*}}
 \frac{i_*-1}{N} +
\frac{\rho_{i_*-1}-1.5}{\rho_{i_*-1}-\rho_{i_*}}
  \frac{i_*}{N}.
\end{equation}
We then have
\begin{equation}
  \mu^I= \frac{1.5-\rho_{i_*}}{\rho_{i_*-1}-\rho_{i_*}}
  \tilde \mu_{i_*-1}
 +\frac{\rho_{i_*-1}-1.5}{\rho_{i_*-1}-\rho_{i_*}}
 \tilde \mu_{i_*} .
\label{muI-formu} 
\end{equation}
Using this formula, we have $\mu^I=7.7 \times 10^{-3}$ for the data
of $\keff=0.5$, and $\mu^I=2.1 \times 10^{-4}$ for the data of
$\keff=1.5$. In Fig. \ref{fig:muI}, $\mu^I$ obtained
by \eqref{muI-formu} are plotted for several values of $\keff$.

Now,  we compare the numerical results with the theoretical
predictions. We developed the theory of the steady state
in the weak noise limit $\Teff \ll 1$ and  the macroscopic
limit $N \gg 1$, with particularly focusing on the two regimes 
$\keff \gg 1$ and $\keff \ll 1$. When $\keff \gg 1$, the chemical potential
at the interface is $\muc=0$, as shown in Appendix \ref{app:con}.
When $\keff \ll 1$, we have the formula
\eqref{muI-final}, where $\muc=0$, $\sigmaL=2.5$, and $\sigmaG=1.5$ were
already determined for the model in the first paragraph of this
section.  $\Xeq$ in the right-side of \eqref{muI-final} is determined
as $\Xeq=1/2$ using \eqref{Xeq-det} with $\bar \rho=1.5$, 
$\rhoLc=2.5$, $\rhoGc=0.5$. By substituting these values into
\eqref{muI-final}, we obtain
\begin{align}
\mu_{X_*}^\inter=\frac{\phi}{6}+O(\phi^2).
\label{muI-neq}
\end{align}
The dotted lines in Fig. \ref{fig:muI} represent the theoretical
predictions $\mu^\inter/\phi=1/6$ for $\keff \ll 1$ and  
$\mu^\inter/\phi=0$ for $\keff \gg 1$. These are consistent
with the numerical result in Fig. \ref{fig:muI}.

It is quite interesting to elucidate the $\keff$ dependence
of $\mu^I$ quantitatively. In particular, one may conjecture
a phase transition at some value of $\keff$ in the limit $N \to \infty$. 
To investigate the validity of this naive conjecture, 
we have to numerically study the asymptotic behavior for
$N \to \infty$, $\Teff \to 0$ and $\phi \to 0$
in more detail. From the theoretical viewpoint, we need to develop
a calculation method for thermodynamic properties of the system
with finite $\keff$.

\section{Concluding remarks}
\label{concluding}

We have derived the variational function determining the steady state for a boundary-driven diffusive system with $\keff \ll 1$.
The result is consistent with global thermodynamics, which is an extended framework of thermodynamics. Before ending the paper, we present a few remarks.


The first remark is on the boundary conditions. It is natural 
to study a system with different boundary conditions leading to the same most
probable profile. As more familiar boundary conditions,
one considers the case that chemical potentials at boundaries are fixed.
However, as far as we attempted, we could not evaluate the
Zubarev-McLennan representation for this case. To study the boundary condition
dependence of the system is a future problem.


Second, in general, fluctuating hydrodynamics is regarded as a mesoscopic
model obtained by coarse-graining microscopic dynamics. Thus, 
it is a significant problem to find relationship between microscopic
dynamics and the discrete fluctuating dynamics.  As the first step
of such studies, parameter values of the model should be determined
from the observation of microscopic dynamics. In particular, it
seems highly challenging to identify the value of $\keff$ from
microscopic models.


Third, as a generalization of the present model, one may consider
a discrete fluctuating hydrodynamics describing liquid-gas
phase coexistence in heat conduction systems. One can numerically
study the model by changing $\keff$. It is interesting
to observe the deviation of the interface temperature from the  equilibrium 
coexistence temperature. Furthermore, following the theoretical method
presented in this paper, we may develop a theory for calculating the
deviation. We conjecture that the deviation formula is equivalent to
that predicted by global thermodynamics.


The most important future work is an experimental observation of non-equilibrium phase coexistence in which metastable states are stable as the influence of a non-equilibrium current. As shown in this paper, the phenomenon is expected to occur in systems described by a discrete fluctuating hydrodynamics. However, it is not obvious whether experimental systems are  described by a discrete fluctuating hydrodynamics. It would be interesting to clarify an experimental condition for realizing a discrete fluctuating hydrodynamics. 


\section*{Acknowledgments}

The authors thank F. Kagawa, K. Saito, and Y. Yamamura for useful discussions
about experiments on non-equilibrium phase coexistence; M. Kobayashi, S. Yukawa, and A. Yoshida for discussions on numerical simulations of non-equilibrium
phase coexistence; and K. Hiura, M. Itami, H. Nakano and Y. Nakayama for discussions on fluctuating hydrodynamics. We also thank B. Derrida, C. Maes,
H. Spohn, and 
H. Tasaki for suggesting us to study boundary-driven diffusive systems.
This work was supported by JSPS KAKENHI Grant Number  
JP22H01144 and JP23K22415.


\vspace{5mm}

\noindent{\small {\bf Data Availability}:  Data sharing  not
    applicable to this article as no datasets were
generated or analysed during the current study.}

\vspace{5mm}

\noindent{\small {\bf Conflict of Interest}:  The authors have no
  financial or proprietary interests in any material discussed
  in this article.
}

\appendix\normalsize
\renewcommand{\theequation}{\Alph{section}.\arabic{equation}}
\setcounter{equation}{0}
\makeatletter
  \def\@seccntformat#1{%
    \@nameuse{@seccnt@prefix@#1}%
    \@nameuse{the#1}%
    \@nameuse{@seccnt@postfix@#1}%
    \@nameuse{@seccnt@afterskip@#1}}
  \def\@seccnt@prefix@section{Appendix }
  \def\@seccnt@postfix@section{:}
  \def\@seccnt@afterskip@section{\ }
  \def\@seccnt@afterskip@subsection{\ }
\makeatother

\section{Derivation of (\ref{canonical})}
\label{app:psi}

In this section, we derive the the stationary distribution
of $\bv{\rho}$ for the equilibrium system with $\phi=0$.
In general, to analyze statistical properties of $\bv{\rho}(t)$
obeying (\ref{model:con}) with (\ref{model:cur}), it is convenient
to use  a new variable $\bv{\psi}=(\psi_i)_{i=1}^N$ for $\bv{\rho}$.
The introduction of the new variable is another purpose of this
section. See Appendix \ref{sec:ZM} for the analysis using  $\bv{\psi}$.

We first define $\psi_i(t)$ by
\begin{equation}
\der{\psi_i(t)}{t}= -j_i(t)
\label{current-psi}
\end{equation}
with $\psi_i(0)=\Lambda \sum_{j=1}^i \red{(\rho_j(0)-\bar \rho)}$
 at $t=0$.
Substituting \eqref{current-psi} into \eqref{model:con} and
integrating it  in time, we  have 
\begin{equation}
  \rho_i(t)= \frac{\psi_i(t)-\psi_{i-1}(t)}{\Lambda}\red{+\bar \rho}
\label{psi-def}
\end{equation}
for any $t$. Substituting \eqref{psi-def} into
${\cal F}(\bv{\rho})$ given by \eqref{Free-form}, we can
define ${\cal F}(\bv{\psi})$ from ${\cal F}(\bv{\rho})$.
Taking the derivative of ${\cal F}(\bv{\psi})$ in $\psi_i$,
we obtain
\begin{align}
  \pder{{\cal F}}{\psi_i} = \red{ \left(\pder{{\cal F}}{\rho_i}-
                             \pder{{\cal F}}{\rho_{i+1}} \right)
                              \frac{1}{\Lambda}}
                          =  \tilde \mu_i- \tilde \mu_{i+1}.
\label{psi-der}                             
\end{align}
Using \eqref{current-psi} and \eqref{psi-der}, we rewrite 
(\ref{model:cur}) as 
\begin{equation}
\frac{d \psi_i}{dt}=
\frac{\sigma(\rho_i^{\rm m})}{\Lambda}
   \left( -\pder{{\cal F}}{\psi_i}- \phi \delta_{i,N} \right)
+  \sqrt{ \frac{2 \sigma(\rho_i^{\rm m}) \Teff}{\Lambda} }\cdot \xi_i
\label{psi-evol}
\end{equation}
with 
\begin{equation}
  \rho_i^{\rm m}= \frac{\psi_{i+1}-\psi_{i-1}}{2 \Lambda}+\bar \rho.
\end{equation}  
Because $\sigma(\rho_i^{\rm m})$ is independent of $\psi_i$,
the multiplication of $\sigma(\rho_i^{\rm m})$ and $\xi_i$
is uniquely determined independently of the multiplication
rule. From (\ref{psi-evol}), we obtain the stationary distribution
of $\bv{\psi}$ for the equilibrium system with $\phi=0$ as
\begin{equation}
{\cal P}_{\eq}(\bv{\psi})=\frac{1}{Z'} \exp
  \left[ -\frac{1}{\Teff} {\cal F}(\bv{\psi}) \right],
\label{canonical-psi}
\end{equation}
where $Z'$ is the normalization constant. 
This gives the stationary distribution
of $\bv{\rho}$ as (\ref{canonical}).

\section{Analysis of the Continuum model}
\label{app:con}

In this section, we analyze the continuum model 
(\ref{continuity}) and (\ref{current}) which corresponds to
the discrete model (\ref{model:con}) and (\ref{model:cur})
in the limit $\keff \to \infty$ and $N \to \infty$ with $L$ and $\kappa$
fixed. Explicitly, we derive the phase
coexistence condition (\ref{n-con}) for the equilibrium
system with $\phi=0$ and the chemical potential at the
interface for the case $\phi >0$.

\subsection{Equilibrium phase coexistence}

Stationary solutions of (\ref{continuity}) and (\ref{current}) with 
$\phi=0$ and $T=0$ satisfy 
\begin{equation}
  \partial_x \var{\calF}{\rho(x)} =0.
\label{ss-sol}
\end{equation}
Using \eqref{Free-form-con},  \eqref{ss-sol} 
is explicitly written as
\begin{equation}
f'(\rho)-\kappa \partial_x^2 \rho =\muc  ,
\label{1st}
\end{equation}
where $\muc$ is a constant in $x$. 
Furthermore, multiplying 
$(\partial_x\rho)$ with (\ref{1st}) and integrating
in $x$, we obtain 
\begin{equation} 
f(\rho)-\frac{\kappa}{2} (\partial_x\rho)^2 -\muc \rho= \pc,
\label{2nd}
\end{equation}
where $\pc$ is also a constant in $x$.

We first consider necessary conditions
under which there is a stationary and spatially inhomogeneous
solution, which we call the phase coexistence solution, because it 
connects two stationary and spatially homogeneous solutions that
represent a liquid phase and a gas phase, respectively. 
To make the argument clear, we take  the limit $L \to \infty$
and we assume $\rho(0) > \rho(\infty)$ for the phase coexistence
solution. Because the phase coexistence solution approaches to the stationary
and spatially homogeneous solutions $\rho(0)$ and $\rho(\infty)$
as $x \to 0$ and $x \to \infty$,  we have
\begin{equation}
  \partial_x \rho(0)=0, \qquad \partial_x^2 \rho(0)=0,
 \qquad \partial_x \rho(\infty)=0, \qquad \partial_x^2 \rho(\infty)=0.
\end{equation}
Therefore, (\ref{1st}) and (\ref{2nd}) lead to
necessary conditions as
\begin{equation}
\mu(\rho(0))=\mu(\rho(\infty))=\muc,
\label{mu-rel}
\end{equation}
and 
\begin{equation}
f(\rho(0)) -\muc \rho(0)= f(\rho(\infty)) -\muc \rho(\infty)=\pc,
\label{p-rel-2}
\end{equation}
which is further rewritten as 
\begin{equation}
\muc= \frac{f(\rho(0))-f(\rho(\infty))}{\rho(0)-\rho(\infty)}. 
\label{p-rel}
\end{equation}
For the function $f(\rho)$ with two local minima, the
conditions \eqref{mu-rel} and \eqref{p-rel} represents
the common tangent line at the special values $\rho=\rhoLc$ and
$\rho=\rhoGc$. We set $\rhoLc > \rhoGc$ without loss of generality.
We thus identify $\rho(0)=\rhoLc$ and $\rho(\infty)=\rhoGc$,
and the values of the constants $\muc$ and $\pc$ are also determined.
Note that (\ref{mu-rel}) and (\ref{p-rel-2}) are regarded as the
conditions giving $\rhoLc$, $\rhoGc$, $\muc$ and $\pc$ by the form
\begin{equation}
\mu(\rhoLc)=\mu(\rhoGc)=\muc, \qquad p(\rhoLc)=p(\rhoGc)=\pc.
\label{p-rel-s}
\end{equation}

Now, suppose that $\bar \rho$ satisfies $\rhoGc < \bar \rho < \rhoLc$,
where $\rhoGc$ and $\rhoLc$ are determined by 
\eqref{p-rel-s}. By setting $\rho(0)=\rhoLc$ and $\rho(\infty)=\rhoGc$,
we solve \eqref{1st} with the determined value of $\muc$.  
We here notice that \eqref{1st} is interpreted as Newton's equation
describing the motion of a point particle under a potential field,
where $\rho$ and $x$ correspond to position and time, respectively.
$\kappa$ is interpreted as the mass, and the potential function $V(\rho)$
is given by $V(\rho) \equiv \muc \rho -f(\rho)$. \eqref{2nd} represents
the energy conservation for the equation of motion. From
\eqref{p-rel-s}, we find that $\rhoLc$ and $\rhoGc$ are local maximal
points with the same potential value. Therefore, the phase coexistence
solution $\rho(x)$ with $\rho(0)=\rhoLc$ and $\rho(\infty)=\rhoGc$ is
given by the heteroclinic orbit connecting the two maximal points with
the same potential value. This result corresponds to the statement
involving (\ref{n-con}) in the main text.

\subsection{Non-equilibrium phase coexistence}

Stationary solutions of (\ref{continuity}) and (\ref{current})
with $\phi > 0$ and $T=0$ satisfy
\begin{equation}
\partial_x   \var{\calF}{\rho(x)} = - \frac{J}{\sigma(\rho(x))} ,
\label{con-J-def}
\end{equation}
which corresponds to  \eqref{v-eq:neq} in the continuum limit 
$\keff \to \infty$ and $N \to \infty$, where $J$ is a constant given by
\begin{equation}
J=\left( \int^{L}_{0} dx~\frac{1}{\sigma( \rho(x))} \right)^{-1},
\end{equation}
which corresponds to \eqref{Jdef-dis} in the continuum limit 
$\keff \to \infty$ and $N \to \infty$.
By analyzing \eqref{con-J-def}, we determine
the value of the chemical potential at the interface.

First, to uniquely identify the interface position, we introduce a
scaled coordinate $\hat x =x/L$. Taking the limit
$L \to \infty$, we find that the interface width in the scaled
coordinate space becomes zero. Thus, the interface position $X$
in the $\hat x$ space is given by the discontinuous point of
$\rho(\hat x)$ in the limit $L \to \infty$. We then
define the chemical potential at the interface by
$\mu(\hat x=X)$. To determine the value of $\mu(\hat x =X)$,
we consider the generalized chemical potential $\tilde \mu(x)$  given
by
\begin{equation}
  \tilde \mu(x) = \mu(\rho(x))-\kappa \partial_x^2 \rho,
\label{tildemu-def}
\end{equation}
which corresponds to \eqref{tilde-mu-2} in the limit $\keff \to \infty$
and $N \to \infty$.
By integrating \eqref{con-J-def} in the range $[x_1,x_2]$, we have
\begin{equation}
  \tilde \mu(x_2)- \tilde \mu(x_1)
  = - J \int_{x_1}^{x_2} dx~ \frac{1}{\sigma(\rho(x))}
\label{1st-ns}
\end{equation}
for any $x_1$ and $x_2$.  Even though $\sigma(\rho(\hat x))$
is discontinuous at $\hat x=X$, the integration in the right-side
of (\ref{1st-ns}) gives a continuous function in $x_2$ and $x_1$. 
Thus, $\tilde \mu(\hat x)$ is a continuous function in the limit
$L \to \infty$. 

Here, the key idea for the determination of $\mu(\hat x=X)$
is the introduction of the generalized pressure
$\tilde p$ satisfying
\begin{equation}
\rho \partial_x \tilde \mu = \partial_x \tilde p .
\label{tilde-p-def}
\end{equation}
We can explicitly derive $\tilde p$ from \eqref{tildemu-def}
and \eqref{tilde-p-def} as
\begin{equation}
  \tilde p=p(\rho)-\kappa \rho \partial_x^2 \rho
  +\frac{\kappa}{2}(\partial_x \rho)^2,
\end{equation}  
which was first obtained by van der Waals \cite{vanPress}. 
By integrating \eqref{tilde-p-def} in the range $[x_1,x_2]$, we have
\begin{equation}
  \tilde p(x_2)- \tilde p(x_1)
  = - J \int_{x_1}^{x_2} dx~ \frac{\rho(x)}{\sigma(\rho(x))}, 
\label{2nd-ns}
\end{equation}
for any $x_1$ and $x_2$. We find 
from \eqref{2nd-ns} that $\tilde p(\hat x)$ is a continuous function
in the limit $L \to \infty$.  
Note that $\tilde p(x)=\pc$ for the
equilibrium system, where $\pc$ is the constant given in \eqref{2nd}.

Now, using the continuity of $\tilde \mu(\hat x)$ and $\tilde p(\hat x)$ at
$\hat x=X$, we can determine the values of 
$\rho(\hat x=X-\epsilon)$ and $\rho(\hat x=X+\epsilon)$ for
small $\epsilon >0$ in the limit $L \to \infty$. Let $\rho_-$ and
$\rho_+$ be $\rho(\hat x=X-\epsilon)$ and $\rho(\hat x=X-\epsilon)$
for $\epsilon \to 0+$ after taking the limit $L \to \infty$. The continuity
of $\tilde \mu(\hat x)$ and $\tilde p(\hat x)$ at $\hat x=X$ leads to
\begin{equation}
\mu(\rho_-)= \mu(\rho_+), \qquad
  p(\rho_-)= p(\rho_+).
\end{equation}
Recalling \eqref{p-rel-s}, we obtain 
\begin{equation}
\rho_-=\rhoLc, \qquad \rho_+= \rhoGc.  
\end{equation}
Therefore, the chemical potential at the interface is equal to $\muc$.
The result is mentioned in the third paragraph
of Sec. \ref{prel-ne} in the main text.

Finally, we present a remark on the singular continuum description
for the case $\keff \to 0$ in Sec. \ref{SCD}. The chemical
potential $\mu(\hat x)$ is continuous at the interface position
$\hat x=X$ where $\rho(\hat x)$ is discontinuous. In this case,
however, $p(\hat x)$ is discontinuous at $\hat x=X$, as shown
by \eqref{pdiff}. That is,
\eqref{tilde-p-def} does not hold at the interface. This is
the most essential difference between the two cases $\keff \gg 1 $
and $\keff \ll 1 $.

\section{Zubarev-McLennan representation}
\label{sec:ZM}

In this section, we derive the Zubarev-McLennan representation
(\ref{ZM}) in Sec.  \ref{sec:zm}. 

We study stochastic processes of $\bv{\psi}=(\psi_i)_{1\le i \le N} $
defined by (\ref{current-psi}).
The time evolution  of $\bv{\psi}$ is
described by (\ref{psi-evol}).
Let $\hat{\bv{\psi}}=(\bv{\psi}_t)_{t=0}^\tau$ be a trajectory 
in the time interval $[0,\tau]$.
The path probability
density ${\cal P}_\path(\hat{\bv{\psi}})$ in the system with $\phi >0$
starting from a density profile sampled from an equilibrium
distribution ${\cal P}_\eq(\bv{\psi}_0)$ is expressed as
\begin{align}
  {\cal P}_\path(\hat{\bv{\psi}})
&=  {\cal P}_\eq(\bv{\psi}_0)  \nonumber \\
  & \times {\rm const} \times\exp\left( - \frac{1}{4\Teff}
    \int_0^\tau dt \sum_{i=1}^N \frac{\Lambda}{\sigma(\rho_i^{\rm m})}
    \left[ \der{\psi_i}{t}+\frac{\sigma(\rho_i^{\rm m})}{\Lambda}
     \left( \pder{\calF}{\psi_i}+\phi \delta_{i,N}\right) \right]^2
     \right) .
\end{align}
For the time-reversed trajectory $\hat{\bv{\psi}}^\dagger
=(\bv{\psi}_{\tau-t})_{t=0}^\tau$ of $\hat{\bv{\psi}}$, we have
\begin{align}
  {\cal P}_\path(\hat{\bv{\psi}}^\dagger)
&=  {\cal P}_\eq(\bv{\psi}_\tau)  \nonumber \\
  & \times {\rm const} \times \exp\left( - \frac{1}{4 \Teff }
    \int_0^\tau dt \sum_{i=1}^N \frac{\Lambda}{\sigma(\rho_i^{\rm m})}
    \left[ -\der{\psi_i}{t}+\frac{\sigma(\rho_i^{\rm m})}{\Lambda}
     \left( \pder{\calF}{\psi_i}+\phi \delta_{i,N}\right) \right]^2
     \right) .
\end{align}
The ratio of the two yields
\begin{align}
  \frac{{\cal P}_\path (\hat{\bv{\psi}})}
       {{\cal P}_\path (\hat{\bv{\psi}}^\dagger)}
         =\exp\left[
           -\frac{1}{ \Teff} \phi\int_0^\tau dt \der{\psi_N(t)}{t} \right],
\label{path-ratio}
\end{align}
where \eqref{canonical-psi} has been  substituted into
${\cal P}_\eq(\bv{\psi}_0)$ and ${\cal P}_\eq(\bv{\psi}_\tau)$.

To simplify the notation, we introduce the accumulated current 
\begin{equation}
  Q_\tau(\hat{\bv{\psi}})\equiv \int_0^\tau dt~ j_N(t) .
\label{Q-def-tau}
\end{equation}
Using \eqref{current-psi} and  \eqref{Q-def-tau}, we rewrite
\eqref{path-ratio} as 
\begin{align}
  \frac{{\cal P}_\path (\hat{\bv{\psi}})}
       {{\cal P}_\path (\hat{\bv{\psi}}^\dagger)}
         =\exp\left[
           \frac{\phi Q_\tau}{ \Teff}  \right].
\label{path-ratio-2}
\end{align}
The distribution of $\bv{\psi}$ at time $t$ is expressed as
\begin{align}
{\cal P}_\tau(\bv{\psi})
& =  \int {\cal D}\hat{\bv{\psi}}~ {\cal P}_\path (\hat{\bv{\psi}})
\delta(\bv{\psi}_\tau-\bv{\psi}) \nonumber \\
& =  \int {\cal D}\hat{\bv{\psi}}^\dagger
\exp\left[ \frac{\phi Q_\tau(\hat{\bv{\psi}})}{ \Teff} \right] 
  {\cal P}_\path (\hat{\bv{\psi}}^\dagger) \delta(\bv{\psi}_\tau-\bv{\psi}),
\label{ptau}  
\end{align}
where we have used ${\cal D}\hat{\bv{\psi}}={\cal D}\hat{\bv{\psi}}^\dagger$
and \eqref{path-ratio-2}. When the path integration variable is transformed, 
the right side of (\ref{ptau}) is rewritten as
\begin{align}
&\int {\cal D}\hat{\bv{\psi}}~
\exp\left[\frac{\phi Q_\tau(\hat{\bv{\psi}}^\dagger)}{ \Teff} \right] 
 {\cal P}_\path (\hat{\bv{\psi}}) \delta(\bv{\psi}_0-\bv{\psi}), \nonumber \\
 =&
 \int {\cal D}\hat{\bv{\psi}}~
\exp\left[-\frac{\phi Q_\tau(\hat{\bv{\psi}})}{ \Teff} \right] 
 {\cal P}_\path (\hat{\bv{\psi}}) \delta(\bv{\psi}_0-\bv{\psi}),
  \label{ptau2}  
\end{align}
where we have used $Q_\tau(\hat{\bv{\psi}}^\dagger)=-Q_\tau(\hat{\bv{\psi}})$.
We thus have the relation
\begin{align}
  {\cal P}_\tau(\bv{\psi})={\cal P}_\eq(\bv{\psi})
  \bra \exp\left[-\frac{\phi Q_\tau}{ \Teff} \right]
  \ket_{\bv{\psi}}.
\end{align}
Taking the limit $\tau \to \infty$, we have
\begin{align}
  {\cal P}_\ss(\bv{\psi})={\cal P}_\eq(\bv{\psi})
  \bra \exp\left[-\frac{\phi Q}{ \Teff} \right]
  \ket_{\bv{\psi}} 
\end{align}
with
\begin{align}
Q= \int_0^\infty dt~ j_N(t).
\end{align}
In the limit $\Teff \to 0$, we estimate
\begin{align}
  \bra \exp\left[-\frac{\phi Q}{ \Teff} \right]
  \ket_{\bv{\psi}}
  \simeq  \exp\left[-\frac{\phi \bra Q\ket_{\bv{\psi}}}
              {\Teff} \right]. 
\end{align}
We then expand $\bra Q_\infty \ket_{\bv{\psi}}$ in $\phi$,
we obtain
\begin{align}
  {\cal P}_\ss(\bv{\psi})={\cal P}_\eq(\bv{\psi})
  \exp\left[-\frac{\phi\bra Q \ket_{\bv{\psi}}^\eq+ O(\phi^2) }{ \Teff} 
    \right] .
\end{align}
This is the Zubarev-McLennan representation of the steady state
distribution.  By using (\ref{psi-def}), we obtain the stationary
distribution of $\bv{\rho}$ as the form (\ref{ZM}).

\section{Derivation of \eqref{Phi0}}
\label{sec:Phi-Phi0}

In \eqref{j0-form}, we consider the decomposition
of $j(0,t)$ into $\Phi(t)$ and $\Phi_0(t)$. In this
section, we calculate $\Phi_0(t)$.

\subsection{Preliminaries for the calculation}

We first note that \eqref{e:DL-DG} and \eqref{e:Yd-Yu}  yield
\begin{align}
\frac{d\hatYd}{dt}=\frac{d\Ys}{dt}+\frac{1}{2}\frac{d|\DL|}{dt}, \qquad
\frac{d\hatYu}{dt}=\frac{d\Ys}{dt}-\frac{1}{2}\frac{d|\DL|}{dt}, \qquad
\frac{d|\DG| }{dt}=-\frac{d|\DL| }{dt} .
\end{align}
From the chain rule of the derivative, we have
\begin{align}
\frac{d}{dt}\left[\frac{|\DL|}{\sigmaL}+\frac{|\DG|}{\sigmaG}\right]^{-1}
&=-\left[\frac{|\DL|}{\sigmaL}+\frac{|\DG|}{\sigmaG}\right]^{-2}\circ 
\left[\frac{1}{\sigmaL}\frac{d|\DL|}{dt}+\frac{1}{\sigmaG}\frac{d|\DG|}{dt}\right]\nonumber\\
&=-\left(\frac{1}{\sigmaL}-\frac{1}{\sigmaG}\right)\left[\frac{|\DL|}{\sigmaL}+\frac{|\DG|}{\sigmaG}\right]^{-2}\circ \frac{d|\DL|}{dt},
\label{e:d1}
\end{align}
and
\begin{align}
&\frac{d}{dt}\left[\frac{1}{\sigmaL}\int_{\DL}dx~\psi+\frac{1}{\sigmaG}\int_{\DG}dx~\psi\right]
=\frac{d}{dt}\left[\frac{1}{\sigmaL}\int_{\hatYu(t)}^{\hatYd(t)}dx~\psi+\frac{1}{\sigmaG}\int_{\hatYd(t)}^{\hatYu(t)+1}dx~\psi\right]\nonumber\\
&=\frac{1}{\sigmaL}\left[\psi(\hatYd(t),t) \circ \frac{d\hatYd}{dt}  -  \psi(\hatYu(t),t) \circ \frac{d\hatYu}{dt}\right]
+\frac{1}{\sigmaG}\left[\psi(\hatYu(t),t) \circ \frac{d\hatYu}{dt}  -  \psi(\hatYd(t),t) \circ \frac{d\hatYd}{dt}\right]\nonumber\\
&\qquad +\frac{1}{\sigmaL}\int_{\DL}dx~ \partial_t \psi  +  \frac{1}{\sigmaG}\int_{\DG}dx~ \partial_t \psi \nonumber\\
&=\left(\frac{1}{\sigmaL}-\frac{1}{\sigmaG}\right)   
\left[
\left(\psi(\hatYd(t),t) -\psi(\hatYu(t),t) \right)\circ \frac{d\Ys}{dt}
+\frac{1}{2}\left(\psi(\hatYd(t),t) +\psi(\hatYu(t),t) \right)\circ \frac{d|\DL|}{dt}
\right]\nonumber\\
&\qquad   +\frac{1}{\sigmaL}\int_{\DL}dx~ \partial_t \psi  +  \frac{1}{\sigmaG}\int_{\DG}dx~ \partial_t \psi .
\label{e:d2}
\end{align}

\subsection{Relation between $\Phi(t)$ and $j(0,t)$}

Noting the two derivatives \eqref{e:d1} and \eqref{e:d2}, we define
\begin{align}
\Phi_1 &\equiv
-\left( \frac{1}{\sigmaL}-\frac{1}{\sigmaG} \right)
\left[
\frac{|\DL|}{\sigmaL}+\frac{|\DG|}{\sigmaG}
\right]^{-2}  
\left[
\frac{1}{\sigmaL}\int_{\DL(t)}dx~  \psi
  +
  \frac{1}{\sigmaG}\int_{\DG(t)}dx~ \psi
\right]
\circ \frac{d|\DL|}{dt},
\label{phi1} \\
\Phi_2 & \equiv
\left( \frac{1}{\sigmaL}-\frac{1}{\sigmaG} \right)
\left[
\frac{|\DL|}{\sigmaL}+\frac{|\DG|}{\sigmaG}
\right]^{-1}  \nonumber \\
& \times \left[
\left(\psi(\hatYd(t),t) -\psi(\hatYu(t),t) \right)\circ \frac{d\Ys}{dt}
+\frac{1}{2}\left(\psi(\hatYd(t),t) +\psi(\hatYu(t),t) \right)\circ \frac{d|\DL|}{dt}.
\right]
\label{phi2}
\end{align}
Then, the definition of $\Phi(t)$ in \eqref{Phi-def} leads to
\begin{align}
\Phi(t)=\Phi_1(t)+\Phi_2(t)+j(0,t)
\end{align}
by using \eqref{j-exp}.  Here, from the piece-wise linear nature of
$\psi(x,t)$, we have
\begin{align}
\frac{1}{\sigmaL}\int_{\DL(t)}dx~  \psi
  +
  \frac{1}{\sigmaG}\int_{\DG(t)}dx~ \psi
=\frac{1}{2}
\left[
\frac{|\DL|}{\sigmaL}+\frac{|\DG|}{\sigmaG}
\right]
\left(\psi(\hatYd(t),t) +\psi(\hatYu(t),t) \right).
\end{align}
See also \eqref{int-psi} for the same equation.
Using this relation, we obtain
\begin{equation}
\Phi_1 +\Phi_2= 
\left( \frac{1}{\sigmaL}-\frac{1}{\sigmaG} \right)
\left[
\frac{|\DL(t)|}{\sigmaL}+\frac{|\DG(t)|}{\sigmaG}
\right]^{-1}  
\left(\psi(\hatYd(t),t)-\psi(\hatYu(t),t)\right)
\circ \frac{d\Ys}{dt}  .
\label{Phi1+Phi2}
\end{equation}
Finally, we note that 
\begin{align}
\psi(\hatYd(t),t)-\psi(\hatYu(t),t)&=\int_{\DL}dx~\partial_x\psi,\nonumber\\
&=|\DL(t)|(\rhoL(t)-\bar\rho).
\end{align}
Substituting this into \eqref{Phi1+Phi2}, we obtain 
\eqref{Phi0} where $\Phi_0\equiv \Phi_1+\Phi_2$.


\end{document}